\documentclass[fleqn,usenatbib]{mnras}

\usepackage{newtxtext,newtxmath}

\usepackage[T1]{fontenc}
\usepackage{ae,aecompl}

\usepackage{graphicx}	

\usepackage[compatibility=false]{caption}
\usepackage{subcaption}

\DeclareMathOperator\erfc{erfc}

\title[General applicability of self-similar winds]{The general applicability of self-similar solutions for thermal disc winds}

\author[A. D. Sellek et al.]{
Andrew D. Sellek,$^{1}$\thanks{E-mail: ads79@cam.ac.uk}
Cathie J. Clarke,$^{1}$ and
Richard A. Booth$^{1,2}$
\\
$^{1}$Institute of Astronomy, University of Cambridge, Madingley Road, Cambridge CB3 0HA, UK\\
$^{2}$Imperial College London, Blackett Laboratory, Prince Consort Road, London SW7 2AZ, UK\\
}

\date{Accepted XXXX. Received YYYY; in original form ZZZZ}

\pubyear{2021}

\begin{document}
\label{firstpage}
\pagerange{\pageref{firstpage}--\pageref{lastpage}}
\maketitle

\begin{abstract}
Thermal disc winds occur in many contexts and may be particularly important to the secular evolution and dispersal of protoplanetary discs heated by high energy radiation from their central star.
In this paper we generalise previous models of self-similar thermal winds - which have self-consistent morphology and variation of flow variables - to the case of launch from an elevated base and to non-isothermal conditions.
These solutions are well-reproduced by hydrodynamic simulations, in which, as in the case of isothermal winds launched from the mid-plane, we find winds launch at the \textit{maximum} Mach number for which the streamline solutions extend to infinity without encountering a singularity.
We explain this behaviour based on the fact that lower Mach number solutions do not fill the spatial domain.
We also show that hydrodynamic simulations reflect the corresponding self-similar models across a range of conditions appropriate to photoevaporating protoplanetary discs, even when gravity, centrifugal forces, or changes in the density gradient mean the problem is not inherently scale free.
Of all the parameters varied, the elevation of the wind base affected the launch velocity and flow morphology most strongly, with temperature gradients causing only minor differences. We explore how launching from an elevated base affects Ne II line profiles from winds, finding it increases (reduces) the full width at half maximum (FWHM) of the line at low (high) inclination to the line of sight compared with models launched from the disc mid-plane and thus weakens the dependence of the FWHM on inclination.

\end{abstract}




\section{Introduction}
Winds originating from accretion discs are thought to account for blueshifted features in a number of astrophysical spectra, such as 
the Low Velocity Component (LVC) of forbidden emission lines such as [\ion{Ne}{II}] and [\ion{O}{I}] in protoplanetary disc line spectra \citep{Hartigan_1995,Pascucci_2009,Rigliaco_2013,Simon_2016,Banzatti_2019,Pascucci_2020}, warm absorbers in AGN spectra \citep[e.g.][]{Mizumoto_2019,Laha_2021,Ganguly_2021} and Fe lines from X-ray binaries \citep[e.g.][]{Begelman_1983,Higginbottom_2020}.
In different environments, various effects may drive or assist the acceleration of the wind: from magnetic tension, to centrifugal forces, to thermal pressure gradients and radiation pressure \citep{Alexander_2014,King_2015,EP_2017}. Understanding the kinematics of these winds is of importance both for interpreting these observations and for assessing their effects both on the accretion disc and their surroundings.

Roughly speaking, a thermal wind may be launched from a disc so long as the thermal energy of heated gas in its upper layers (which is converted to kinetic energy by pressure gradients) is sufficient to overcome the gravitational potential of the star, leading to unbound material. For a given sound speed $c_S$, this is possible outside of the gravitational radius $r_G$ \citep{Shu_1993,Hollenbach_1994}:
\begin{equation}
    r_G = \frac{GM_*}{c_S^2} \approx 9~\mathrm{au} \frac{M_*}{M_{\sun}} \left(\frac{c_S}{10~\mathrm{km~s^{-1}}}\right)^{-2}
    \label{eq:r_G}.
\end{equation}
In practice, pressure gradients mean that winds are also accelerated inside $r_G$ \citep{Font_2004,Clarke_2016}, with $r_{\rm crit} \approx 0.2 r_G$ taken as the typical limit \citep{Alexander_2014}.
By comparison, the typical gas radii of protoplanetary discs are much larger, $\gtrsim 100~\mathrm{au}$ \citep{Barenfeld_2017,Ansdell_2018}.

Within the context of protoplanetary discs, a thermally-driven disc wind, heated by the central star, is thought to be one of the key mechanisms to disperse the discs in a process known as "Internal Photoevaporation"\footnote{In contrast, "External Photoevaporation" involves heating by primarily Far Ultraviolet radiation from nearby O and/or B type stars and is most relevant in large, dense, star-forming regions.}. Photoevaporative winds are often favoured since discs are largely thought to disperse from the inside out \citep{Koepferl_2013}, but must do so rapidly since the fraction of so-called Transition Discs that appear to be undergoing clearing is small \citep[$\leq10$ per cent][]{Andrews_2005}. The rapid clearing possible under direct irradiation could account for this observed behaviour in many discs \citep[e.g.][]{Owen_2011,Picogna_2019}, though competing ideas include planet-disc interactions. Due to their potential importance, and the abundance of material beyond $r_G$ where thermal forces can dominate, in this work we focus on thermal winds driven from large radii from discs with properties appropriate to protoplanetary discs.

The primary radiation that is responsible for heating internally driven winds is still a matter of debate \citep[e.g.][]{Alexander_2014,EP_2017}, with different bands having distinct effects on the ionisation state, density, temperature and velocity of the outflow and hence its mass loss rate.
For example, the earliest models considered heating due to the ionising extreme ultraviolet (EUV) radiation \citep{Shu_1993,Hollenbach_1994}. These typically create fully ionised, low density winds that are roughly isothermal at $10^4~\mathrm{K}$ (i.e. sound speeds of $10~\mathrm{km~s^{-1}}$).

Later models consider X-rays to be the predominant heating mechanism \citep{Owen_2010,Owen_2011,Owen_2012,Picogna_2019}. X-rays lead to only partially ionised flows (of much higher density than the EUV winds) in which EUV photons would be absorbed at small radii \citep{Ercolano_2010,Owen_2012} before reaching the bulk of the wind. The deeper penetration of the X-rays allows flows from larger radii, so while the result is a cooler, and hence somewhat slower, wind (with sound speeds of $c_S\approx3-5~\mathrm{km~s^{-1}}$) the mass loss rates can be considerably higher \citep{Ercolano_2009}.

Not only does the temperature of the wind depend on various heating mechanisms, but also how they are offset by cooling.
Using a modest chemical network, \citet{Wang_2017} found that with different cooling mechanisms, X-rays were unable to heat the gas sufficiently and hence they recovered an EUV driven wind with lower mass loss rates.
In this scenario the X-rays would act more like the non-ionising far ultraviolet (FUV) radiation  which may chiefly aid mass loss by heating a layer below the wind \citep{Owen_2012} and puffing up the wind base so that it intercepts more of the radiation that is capable of heating the flow to above the escape velocity \citep{Wang_2017}.
The FUV itself is more challenging to include due to the complicated photochemistry of the photodissociation regions so the first models to treat it used very simplified models for the hydrodynamics of the wind \citep{Gorti_2009a}. Due to lower temperatures to which the FUV heats the gas, on its own it is only effective at driving a flow at very large radii, although at suitably high metallicities, \citet{Nakatani_2018a} found that the FUV could drive a substantial neutral flow.

In order to understand the true significance and viability of photoevaporative winds, it is important to attempt to disentangle these effects using observational constraints. 
The mass loss rates in winds are hard to measure directly, but
constraints from disc demographics imply relatively low rates are more typical
\citep{Alexander_2014,Somigliana_2020,Sellek_2020b}. However, demographics have not been able to discriminate strongly between the different heating mechanisms due to uncertainties in the illuminating flux.

A better probe is to measure the velocities of the outflowing gas using the aforementioned optical and infrared forbidden lines. In particular, the [\ion{Ne}{II}] $12.81~\mu\mathrm{m}$ and [\ion{O}{I}] 6300 \r{A} are potential tracers of a wind. They show blueshifted profiles which may typically consist of High Velocity and Low Velocity Components (HVC and LVC respectively) \citep{Hartigan_1995} where the former is usually attributed to a magnetohydrodynamic jet. Modelling of the blueshifts and widths of these components has been used to constrain the origin and heating mechanism of the flows.

The [\ion{O}{I}] 6300 \r{A} LVC has been suggested to trace a disc wind \citep[e.g.][]{Hartigan_1995}. \citet{Font_2004} made the first attempt to model this emission as a thermal wind: while several lines from \textit{ionised} species detected by \citet{Hartigan_1995} could be explained by an EUV driven wind, the \textit{neutral} [\ion{O}{I}] 6300 \r{A} was not consistent with the observed line luminosities. \citet{Ercolano_2016} suggested that this blueshifted [\ion{O}{I}] emission is instead a `smoking gun' of X-ray photoevaporation, though found it was not suitable for measuring the mass-loss rates as the size of the emission region is also determined by the contribution of EUV to the heating.
More recent observations suggest that two components make up the LVC \citep{Simon_2016} - a broad component (BC) and a narrow component (NC) - the former of which is too broad to trace a thermal wind and more likely probes a magnetically driven wind at small radii. 
\citet{Banzatti_2019} further argued that since the properties of the NC were found to correlate with those of the BC, then both components of the LVC should have a magnetic origin, though \citet{Weber_2020} suggested a mechanism by which the NC could still result from a thermally driven wind.

The [\ion{Ne}{II}] $12.81~\mu\mathrm{m}$ emission has been successfully explained by a thermal wind \citep{Alexander_2008,Pascucci_2009,Ercolano_2010,Pascucci_2011,Alexander_2014}. Most recently, \citet{Pascucci_2020} showed that the [\ion{Ne}{II}] LVC was more strongly present in discs with larger inner cavities (in contrast to the [\ion{O}{I}] which diminishes as the inner disc clears) and argued that this points to an origin outside the cavity. This suggests a picture where photoevaporative winds are particularly relevant at large radii in discs with cavities undergoing clearing - whose line profiles demand winds driven from well outside $r_G$ - with full discs possessing an inner magnetohydrodynamic wind that blocks the ionising radiation needed to drive a thermal wind at large radii.

Modelling these observational diagnostics has typically required hydrodynamic simulations to self-consistently calculate the thermal structure and generate the streamline morphology \citep[e.g.][]{Font_2004,Ercolano_2010,Picogna_2019} because in general no analytic solution exists.
However including both radiative transfer and hydrodynamics makes such simulations expensive (particularly if multiple parameters are to be studied) and consequently somewhat limited in their domain, which can compromise the line profiles derived from such simulations.
Consequently, in parallel with efforts to improve the verisimilitude of wind simulations \citep[focusing for example on detailed microphysics surrounding disc photochemistry, ionization, and magnetically driven winds e.g.][]{Wang_2019,Grassi_2020} we take an alternative approach. In this paper we explore the large scale kinematics and morphology of thermally driven winds in cases where the wind temperature structure and the density profile at the flow base are simply prescribed (with dependencies motivated by the results of radiation hydrodynamic simulations) to investigate to what extent we can separate these kinematics and morphology from the microphysics.

The simplest approach is to model winds as isothermal, which is a decent first approximation as the temperature gradients found in simulations tend to be small \citep[e.g.][]{Nakatani_2018a,Picogna_2019}. This approach enabled \citet{Clarke_2016} to develop a self-similar model for isothermal winds in which the streamline morphology is set by a balance between pressure gradients and inertial forces from the streamline curvature. They argued that gravitational and centrifugal effects are subdominant since they cancel at the base of the wind and decline more rapidly along the streamlines (as $1/r^2$ and $1/r^3$ respectively) than the pressure or inertial forces.
They thus obtained self-similar solutions for power law base density profiles in which the wind launches at a constant velocity $u_{\rm b}$;
for a given density profile, while there were a range of self-consistent solutions to their modified `de Laval nozzle' problem, there was a maximum Mach number $\mathcal{M}_{\rm b, max}$ for which the solution could avoid a singularity in the equation at some point along the streamline and thus remain smooth.
When these solutions were benchmarked against scale-free 2D hydrodynamic simulations, they found that the solution adopted by the wind was in good agreement with the maximal allowed solution launched at $\mathcal{M}_{\rm b, max}$.
Moreover, even when they reintroduced gravity and rotation, this solution was well-recovered at radii $r \gg r_G$ and was even a good description at radii as low as $0.5~r_G$, so long as the radius of curvature of the streamlines was small, thereby justifying neglecting gravity and rotation.
Thus \citet{Clarke_2016} concluded that in general one should expect thermal winds to launch at $\mathcal{M}_{\rm b, max}$.

These self-similar solutions enabled \citet{Ballabio_2020} to model the blueshifted [\ion{O}{I}] emission lines observed by \citet{Banzatti_2019}, as well as literature data for [\ion{Ne}{II}].
They showed that the blueshifts were most sensitive to the sound speed; while the [\ion{Ne}{II}] preferred a fast wind with $c_S\sim10~\mathrm{km~s^{-1}}$ as appropriate to an EUV wind, the [\ion{O}{I}] was more consistent with a slower wind with $c_S\sim3-5~\mathrm{km~s^{-1}}$.
On the other hand the full width half maxima of the [\ion{Ne}{II}] lines also favoured a high sound speed whereas for [\ion{O}{I}] a larger sound speed was generally better but did not reproduce the widths well. In all this implies potentially different origins for the [\ion{Ne}{II}] and [\ion{O}{I}] emission in agreement with \citet{Pascucci_2020}.

The solutions of \citet{Clarke_2016} are, however, limited to power law winds launched perpendicularly from the midplane, and in an isothermal atmosphere. If self-similar solutions are to be used to approximate the results of radiation hydrodynamics simulations and accurately interpret observational data, the results of \citet{Clarke_2016} must be tested with these four assumptions relaxed.
The first two of these - midplane launches and perpendicular launches - are needed as in general simulations such as those by \citet{Wang_2017,Picogna_2019} show streamlines originating from elevated bases, sometimes at less than right angles.
Moreover, as discussed, winds would not be expected to remain perfectly isothermal.
Finally the restriction to power laws in density neglects the fact that a single power law can only apply over a limited range - for example \citet{Hollenbach_1994} propose two different power law regimes either side of $r_G$.

In this work we thus seek to further understand the applicability of the self-similar solutions to modelling thermal disc winds.
In section \ref{sec:mathematical_extension} we summarise the hydrodynamics behind the winds and outline our extension of self-similar solutions to winds launched from elevated bases, non-perpendicularly, and with scale-free temperature profiles. In section \ref{sec:numerical_solutions} we calculate maximum launch velocities using our generalised solution\footnote{We have made public the code we use to do so such that others may easily use the solutions as part of their own modelling: \href{https://github.com/AndrewSellek/SelfSimilarThermalWinds}{https://github.com/AndrewSellek/SelfSimilarThermalWinds}.}, showing that for realistic temperature variations, the effects on the streamline morphology and launch Mach number are negligible.
In section \ref{sec:ss_hydro} we show that in all of these cases, the scale-free hydrodynamic simulations still predict launch velocities corresponding to the maximum values predicted by the self-similar solutions and we provide an argument as to why this is based on these solutions being space-filling in Section \ref{sec:domainfilling}.
We reintroduce gravity and centrifugal forces in section \ref{sec:nss_hydro} and in section \ref{sec:dbl_pow} we investigate the effects of imposing a double power law in density of the form used by \citet{Font_2004} at the base.
We discuss the implications of our results - for both the interpretation of hydrodynamic simulations and application to other problems - in section \ref{sec:discussion}, before summarising our conclusions in section \ref{sec:conclusions}.

\section{Hydrodynamics of Wind Launching Solutions}
\label{sec:mathematical_extension}
We first describe the forces and fluid equations that govern the flow in the self-similar assumption, before going to to describe how we extend the treatment of \citet{Clarke_2016} to winds that are a) launched from elevated bases\footnote{\citet{Hutchison_2021} also performed such a calculation for their specific case of interest.}, b) launched non-perpendicularly to their launch plane, c) subject to scale-free temperature profiles.

The momentum equation equation describing the forces on the gas in either the wind or the underlying disc may be written in spherical polar coordinates (radius $r$) in terms of the density $\rho$, pressure $P$ and gravitational force from the central star $-\frac{GM_*}{r^2}$ as 
\begin{equation}
    \frac{\partial \vec{u}}{\partial t} + \vec{u}\cdot\nabla\vec{u} = -\frac{1}{\rho} \nabla P - \frac{G M_*}{r^2} \hat{r}
    \label{eq:momentum_equation}
    .
\end{equation}

Rather than balancing forces in coordinates set by the global geometry, we follow \citet{Clarke_2016} who demonstrated that equation \ref{eq:momentum_equation} may be written in a local coordinate system, which in the steady state, when resolved perpendicular to the streamlines (in the direction with unit vector $\hat{l}$), gives 
\begin{equation}
    \frac{u^2}{R_{\rm eff}} = \frac{1}{\rho} \hat{l} \cdot \nabla P
    ,
    \label{eq:eom_perp}
\end{equation}
where $R_{\rm eff}$ is the radius of curvature.
Thus in the self-similar wind model, the pressure gradients normal to the streamlines balance an effective inertial/centrifugal force due to the streamline curvature.

In writing equation \ref{eq:eom_perp}, two important forces have been neglected: centrifugal force due to rotation in the $\phi$ direction (which forms part of the inertial term $\vec{u}\cdot\nabla\vec{u}$)
and gravity.
In the underlying disc at the midplane - which is cold and thin - these two forces balance, since the pressure gradient is subdominant (by a factor $(H/R)^2$).
The material that supplies the wind flows vertically through the cold, thin, disc and eventually passes through the wind base, where it is heated and strongly accelerated. The gravitational and centrifugal forces are barely changed compared to their midplane values, but once in the wind region, which is much hotter than the underlying disc, the pressure gradient has greatly increased.
Since all quantities in the wind solution vary over a length scale of order r or less, the magnitude of the acceleration associated with the pressure gradient, $\frac{1}{\rho} \frac{\partial P}{\partial r} \approx \frac{c_S^2}{r}$, exceeds gravity and centrifugal force at the wind base, so long as $\frac{c_S^2}{r_{\rm b}} > \frac{GM_*}{r_{\rm b}^2}$. I.e. so long as $r>r_G$ (as defined in equation \ref{eq:r_G}) we can make this approximation. Physically, this represents a regime where the thermal energy exceeds the gravitational energy and hence the material is unbound.
As one moves to larger radii, the gravitational and centrifugal terms decline much faster (as $1/r^2$ and $1/r^3$ respectively) than the pressure gradient ($1/r$ in an isothermal disc) and thus the approximation is strengthened.

Since the velocities are on the order of $c_s$, the curvature term also dominates over gravity (and centrifugal force)
if $r > \frac{R_{\rm eff}}{r} r_G$ \citep{Clarke_2016}. The solutions are typically sufficiently curved near the base that $R_{\rm eff} < r \approx r_{\rm b}$,
making the approximation reasonable even for winds launched from $r_{\rm b}$ somewhat inside $r_G$.

The self-similar solutions by construction do not therefore include the cold disc, however one can separately compute the velocity structure matching the mass and momentum flux across the wind base implied by the self-similar wind solutions \citep[see e.g.][]{Hutchison_2021}. In practice, this has limited effect on the vertical structure of the underlying disc, which is in a state of near-hydrostatic equilibrium, so we are not concerned with it in this paper.

The key conservations in this problem are momentum and mass along the streamlines.
Momentum conservation gives us the following relationship between the velocity and pressure gradients (which can equivalently be thought of as the conservation of the Bernoulli function)
along the streamline:
\begin{equation}
    u \frac{\partial u}{\partial s} + \frac{1}{\rho} \frac{\partial P}{\partial s} = u \frac{\partial u}{\partial s} + c_S^2 \frac{\partial\ln(\rho)}{\partial s} + \frac{\partial c_S^2}{\partial s} = 0
    ,
    \label{eq:Bernoulli_general_P}
\end{equation}
where $\frac{\partial}{\partial s}$ is the directional derivative along the streamline and the second form is achieved by expanding in terms of the density and isothermal sound speed $c_S^2 = \left(\frac{\partial P}{\partial \rho}\right)_{T} = \frac{\mathcal{R}}{\mu} T$.

Meanwhile mass conservation implies a constant mass flux
\begin{equation}
    \dot{M} = \rho u A
    ,
    \label{eq:masscons_mdot}
\end{equation}
where $A$ is the area of a streamline bundle. Hence by differentiating and using equation \ref{eq:Bernoulli_general_P} we may write
\begin{align}
    \frac{\partial\ln(A)}{\partial s} &= - \frac{\partial\ln(\rho)}{\partial s} - \frac{\partial\ln(u)}{\partial s} \\
    &= \left( \frac{u^2}{c_s^2} -1 \right) \frac{\partial\ln(u)}{\partial s} + \frac{\partial\ln(c_S^2)}{\partial s}.
    \label{eq:nozzle}
\end{align}
This demonstrates a nozzle effect where the wind is accelerated so long as we have a converging-diverging flow. In the isothermal case the sonic surface coincides with a minimum in the area of the streamline bundle; temperature gradients can offset this slightly.

In Appendix \ref{sec:a1}, we provide a more detailed description of how the pressure and temperature gradients
are calculated in our self-similar model, and how the radius of curvature, $R_{\rm eff}$, may be evaluated once the rate of change of area is known. Combining Equations \ref{eq:eom}-\ref{eq:Reff} and \ref{eq:dlnA_general} representing these gives the single ordinary differential equation given in Section \ref{sec:general_equations}.

\subsection{Description of the Wind Base}
In the approximation described above, since gravity is neglected, then $r_G$ cannot enter the solution - instead, the sole length scale is the radius at the base of each streamline, $r_b$, and so our solutions will be self-similar. This means any quantities with dimensions of length must scale linearly with $r_{\rm b}$, for example the spherical radius, $r$, scales as $r = r_b \tilde{r}(\tilde{s})$ (where $\tilde{s}$ is the normalised arc length along the streamline).
Similarly, since the problem we solve is a initial value problem, other quantities, such as the density and velocity, are most sensibly expressed in terms of their values at the base i.e. $\rho = \rho_{\rm b} \tilde{\rho}(\tilde{s})$ and $u = u_{\rm b} \tilde{u}(\tilde{s})$ respectively.

\citet{Clarke_2016} showed that for the \textit{globally isothermal} case to display self-similarity, $u_{\rm b}$ must be the same for all streamlines. In this work we argue that more generally, when the sound speed (temperature) varies along the wind base, the \textit{Mach number at the base} $\mathcal{M}_{\rm b} = u_{\rm b}/c_S$ must be the same for all streamlines (see Appendix \ref{sec:a1}).

\subsubsection{Geometry}
To describe the wind geometry at the base, we first define the angle $\phi$ as the angle between a point on a streamline and the midplane i.e.
\begin{equation}
    \tan(\phi) = \frac{z(\tilde{s},r_b)}{R(\tilde{s},r_b)} = \frac{\tilde{z}(\tilde{s})}{\tilde{R}(\tilde{s})},
    \label{eq:def_phi}
\end{equation}
where $R$ and $z$ are the radial and vertical coordinates in cylindrical polar coordinates and $\tilde{R}$, $\tilde{z}$ are the equivalents normalised to $r_{\rm b}$\footnote{Note that therefore from equation \ref{eq:def_phi}, $\phi = \phi(\tilde{s})$ so the solution at a given normalised arc length along the streamline could equally be thought of as the solution at given angle from the midplane.}.
$\phi_{\rm b}$ is then the elevation of the wind base.

Similarly, we define the angle between a streamline tangent and the midplane as
\begin{equation}
    \tan(\theta) = \frac{dz}{dR} = \frac{d\tilde{z}}{d\tilde{R}},
\end{equation}
and for ease further define the angle $\chi$ as
\begin{equation}
    \chi = \theta - \phi,
\end{equation}
such that $\chi_{\rm b}=\theta_{\rm b}-\phi_{\rm b}$ represents the angle with which the wind launches relative to the plane from which it launches.

While \citet{Clarke_2016} assumed $\phi_{\rm b}=0$ and $\chi_{\rm b}=\theta_{\rm b}=\pi/2$, in Section \ref{sec:numerical_solutions} we explore the parameter space $0^\circ \leq \phi_{\rm b} \leq 72 ^\circ$ and $0.25\pi \leq \chi_{\rm b} \leq 0.5\pi$ in order to show how the launch Mach numbers $\mathcal{M}_{\rm b}$ depend on these parameters. For most of the paper, however, we focus on an elevation of $\phi_{\rm b}=36^\circ$ - which typical of photoevaporation models \citep{Wang_2017,Picogna_2019} - and a launch angle of $\chi_{\rm b} = \pi/2$ since simulations show material reaching the base with very low velocities, and only being accelerated perpendicularly (see further discussion in section \ref{sec:nss_hydro}).

\subsubsection{Imposed Profiles of Flow Variables}
In order to seek a self-similar solution, we require that the density at the base of the wind $\rho_b$ is a power law (which has no characteristic scale) in base radius $r_b$: 
\begin{equation}
    \rho_b = \rho_0 \left(\frac{r_b}{r_0}\right)^{-b}.
    \label{eq:rhob}
\end{equation}
where $b$ is simply the power law index for which \citet{Clarke_2016} consider values in the range $0.5-2$. We focus throughout much of this paper on $b=1.5$ since this most closely resembles the density gradient at the base in \citet{Picogna_2019}, as well as the density profile for $r<r_G$ found by \citet{Hollenbach_1994}. This value also better reproduces the [\ion{Ne}{II}] line luminosity \citet{Pascucci_2011} measured for TW Hya \citep{Ballabio_2020}.
Note that since $\tilde{s}$ and $\phi$ are interchangeable, then at any fixed elevation $\rho$ and $u$ (or $\mathcal{M}$) should scale in the same way as at the base, e.g. $\rho \propto r^{-b}$.

We impose a fixed temperature $T$ at each location, i.e. use a locally isothermal equation of state $P= \frac{\mathcal{R}}{\mu} \rho T$. \footnote{Note that this condition results from the balance of heating and cooling at each location, and \textit{not} from the wind material being adiabatic with $\gamma\approx1$, which would instead imply material kept the temperature at its base.} For convenience, we express this temperature structure in terms of the isothermal sound speed; the sound speed profile must also be scale free, meaning it can be written in a separable form in terms of $r$ and $\phi$, where the dependence on $r$ is that of a power law: 
\begin{align}
    c_S^2(r,\phi) &= c_{S,\rm b}^2(r_{\rm b}) \tilde{c}_S^2(\tilde{r},\phi) = c_{S,\rm b}^2(r_{\rm b}) \tilde{r}^{-\tau} \mathcal{C}(\phi)
    \label{eq:cs_general}.
\end{align}
where $c_{S,\rm b}^2(r_{\rm b}) \propto r_{\rm b}^{-\tau}$ and
the angular dependence is normalised such that $\mathcal{C}(\phi_b)=1$. $\tau$ is defined (analogously to $b$) as the power law slope of the temperature profile.
For \textit{disc} temperatures, commonly used profiles have $\tau=0.5$ assuming a simple black-body equilibrium dust temperature, or a slightly lower $\tau\approx0.43$ if one includes the effects of the disc flaring \citep{Chiang_1997}. While the heating mechanisms are very different in the \textit{wind}, the principles of geometric dilution \citep{Owen_2012} and disc flaring \citep{Wang_2017} are still important and simulations suggest similarly modest outward temperature gradients \citep[e.g. $0.28<\tau<0.4$ depending on the metallicity,][]{Nakatani_2018a}. We thus investigate fiducial non-isothermal cases with $\tau=0.25$ and $\tau=0.5$.

\subsection{General Equations for Self-Similar Winds}
\label{sec:general_equations}
In  Appendix \ref{sec:a1} we solve the above problem
to derive an equation relating the velocity gradient to the cylindrical coordinates of a point on the streamline ($\tilde{R}$ and $\tilde{z}$), the direction of the streamline tangent $\cot(\theta)=\tilde{R}'=\frac{d\tilde{R}}{d\tilde{z}}$ and the velocity $\tilde{u}$.

\begin{align}
    \frac{{\rm d}\tilde{u}}{{\rm d}\tilde{z}} &= \frac{g_1+g_2}{f_1+f_2}
    \label{eq:odes_0} \\
    f_1 &= - \mathcal{M}_b^2 \tilde{u} \left(\mathcal{M}_{\rm b}^2 \frac{\tilde{u}^2}{\tilde{c}_S^2} - 1 \right)
    \frac{(\tilde{R}-\tilde{z}\tilde{R}')}{(1+\tilde{R}'^2)^{1/2}(\tilde{R}\tilde{R}'+\tilde{z})}
    \label{eq:odes_f1}\\
    f_2 &= \mathcal{M}_b^2 \tilde{u} \frac{\tilde{R}\tilde{R}'+\tilde{z}}{(1+\tilde{R}'^2)^{1/2}(\tilde{R}-\tilde{z}\tilde{R}')}
    \label{eq:odes_f2}\\
    g_1 &= \left(b + \tau \right) \frac{(1+\tilde{R}'^2)^{1/2}  }{(\tilde{R}-\tilde{z}\tilde{R}')} \tilde{c}_{S}^2
    \label{eq:odes_g1} \\
    \begin{split}
    g_2 &= - \mathcal{M}_b^2 \tilde{u}^2
    \frac{(\tilde{R}-\tilde{z}\tilde{R}')}{(1+\tilde{R}'^2)^{1/2}} \\
    & \left( \frac{\tilde{R}'}{\tilde{R}(\tilde{R}\tilde{R}'+\tilde{z})}
    + \frac{1}{(\tilde{R}^2+\tilde{z}^2)} \left( \tau
    - \frac{\partial \ln(\mathcal{C})}{\partial \phi}
    \frac{(\tilde{R}-\tilde{z}\tilde{R}')}{(\tilde{R}\tilde{R}'+\tilde{z})}
    \right) \right)
    \end{split}
    \label{eq:odes_g2}
    .
\end{align}
We will refer to $f_2$ and $g_1$ as pressure related terms and $f_1$ and $g_2$ as curvature related terms since they arise from the right and left hand sides of equation \ref{eq:eom} respectively.
Compared to \cite{Clarke_2016}, $f_2$ is unchanged as it represents the change in pressure due to the Bernoulli effect, which is unchanged by geometry or temperature gradients. $g_1$ represents the radial pressure gradient and hence is affected by the additional radial temperature gradient $\tau$ and scales with the sound speed normalised to the base $\tilde{c}_S^2$. 
$f_1$ takes on a very different form - this is because we calculated $\tilde{A}'$ in a different way to \citet{Clarke_2016}\footnote{In that work they used $\tilde{A}(\tilde{u})$ to calculate $\tilde{A}'$, but this can only be solved explicitly in the isothermal case (by using the conservation of the Bernoulli function). We have checked that the alternate form of the term presented here recovers the same solution in the isothermal case}. Finally $g_2$ picks up an additional terms due to the variation of temperature with both radius and latitude.

\section{Numerical Solution}
\label{sec:numerical_solutions}
The maximum launch Mach numbers $\mathcal{M}_{\rm b,max}$ were calculated by \citet{Clarke_2016} by numerically solving their equivalents to equations \ref{eq:odes_0} to \ref{eq:odes_g2} using an Euler method, iterating to find solutions which avoided $f=f_1+f_2\to0$. We apply the same method to investigate more general geometries for isothermal winds, and the effects of temperature profiles that follow radial power laws.

At each point $\tilde{z}$ along the streamline, we track $\tilde{u}$, $\tilde{R}$ and $\tilde{R}'$.
The velocity gradients are calculated using equation \ref{eq:odes_0}.
Then we can find the gradient of the area of a streamline bundle using equation \ref{eq:dlnA_general} and hence $\tilde{R}''$ from equation \ref{eq:R_dlbprime}.
Finally, $\tilde{u}$, $\tilde{R}$ and $\tilde{R}'$ can be advanced to locate the next point on the streamline and its velocity:
\begin{align}
    \tilde{u}_{i+1}     &=  \tilde{u}_{i} + \tilde{u}'_{i} \Delta \tilde{z} \\
    \tilde{R}_{i+1}     &= \tilde{R}_{i} + \tilde{R}'_{i} \Delta \tilde{z} + \frac{1}{2}  \tilde{R}''_{i} (\Delta \tilde{z})^2 \\
    \tilde{R}'_{i+1}    &= \tilde{R}'_{i} + \tilde{R}''_{i} \Delta \tilde{z}
\end{align}

\subsection{More General Geometries}
We first consider isothermal winds, for which we expect that the maximum launch Mach numbers can now be a function of the angles $\phi_{\rm b}$ and $\chi_{\rm b}$ as well as the density gradient $b$.
Hence we proceed to solve our revised differential equation across this 3-dimensional parameter space, finding $\mathcal{M}_{\rm b,max}(b,\phi_{\rm b},\chi_{\rm b})$ for a regular grid covering density power law slopes $0.5 \leq b \leq 1.75$, elevation angles $0^\circ \leq \phi_{\rm b} \leq 72 ^\circ$ and launch angles $0.25\pi \leq \chi_{\rm b} \leq 0.5\pi$.

\begin{figure*}
    \centering
    \includegraphics[width=0.9\linewidth]{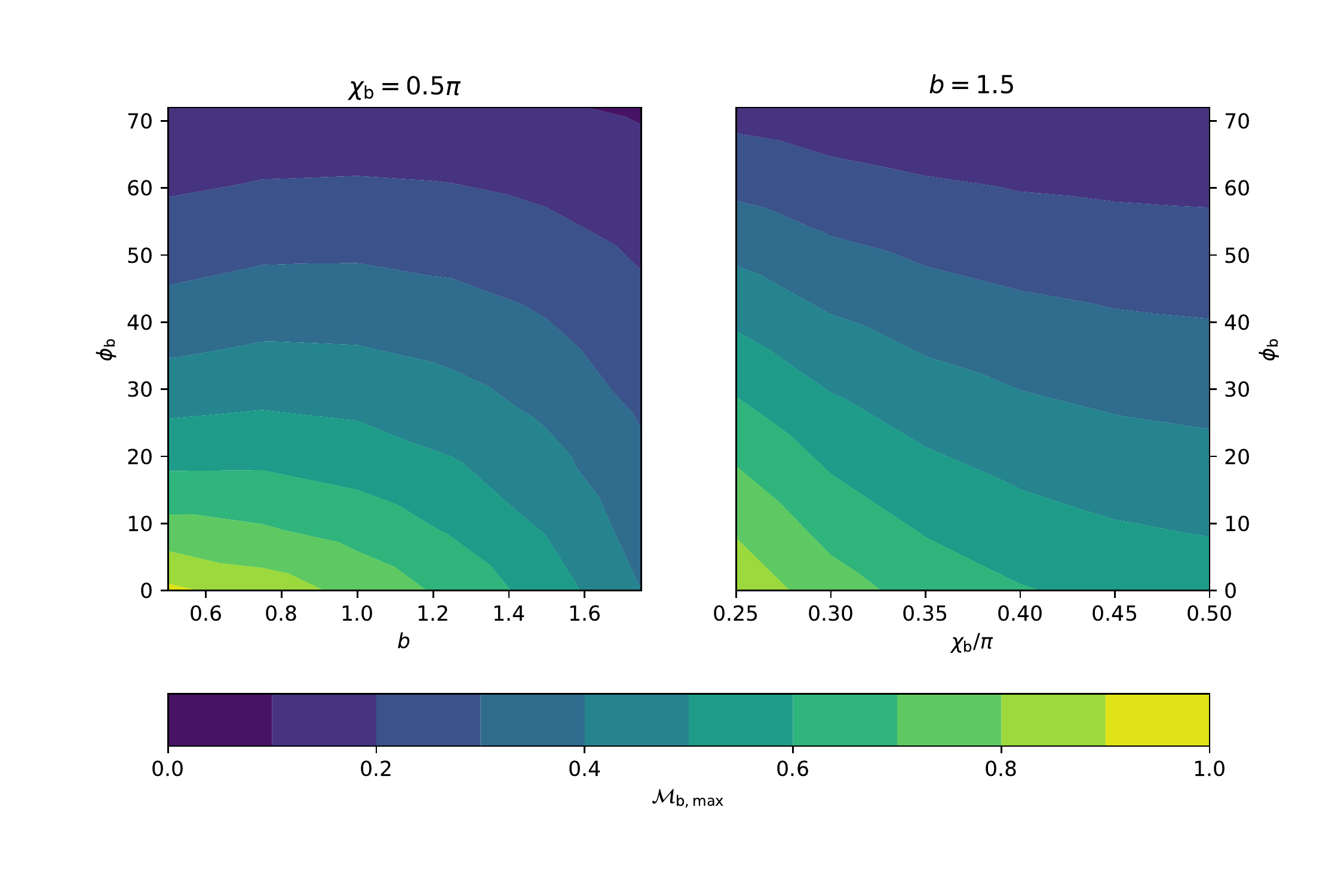}
    \caption{The maximum allowable Mach number at the base $\mathcal{M}_{\rm b, max}$ as a function of the density power law slope $b$, the elevation angle of the wind base $\phi_{\rm b}$ (in degrees) and the angle with which the wind is launched relative to this plane $\chi_{\rm b}$ (in radians). The left-hand panel shows the dependence on $b$ and $\phi_{\rm b}$ for perpendicularly-launched winds, while the right-hand panel shows the dependence on the two angles for a fixed density profile with $b=1.5$. The lighter, yellow colours at low $b$, low $\phi_{\rm b}$ and low $\chi_{\rm b}$ represent faster winds, while the darker blue colours represent slower velocities.}
    \label{fig:Mb_contours}
\end{figure*}

For each model we integrate out to $\tilde{z}=1000$; so long as we do not encounter $f\to0$, we then increase the Mach number by $0.1$ until such a singularity is encountered. Once this scenario arises, we then return to highest safe value and repeat, first increasing $\mathcal{M}_{\rm b}$ by $0.01$ and then repeating a third time increasing by $0.001$ such that $\mathcal{M}_{\rm b}$ is found to 3 decimal places.
The results are presented as the contour plots in Figure \ref{fig:Mb_contours}.

The trends in Figure \ref{fig:Mb_contours} can largely be understood by considering the relationship between $\mathcal{M}_{\rm b}$ and the (radius of) curvature at the base.
Firstly, as found by \citet{Clarke_2016}, the Mach number at the base is generally a decreasing function of the density power law slope $b$. This is because a stronger pressure gradient provides a stronger force to push the streamlines over, meaning that they curve more strongly i.e. with a smaller radius of curvature; more strongly curved winds are associated with a slower velocity.
Moreover, as we increase $\phi_{\rm b}$, there is a strong decrease in $\mathcal{M}_{\rm b,max}$. This is because the winds have to turn to become radial and outward flowing in a tighter space, meaning they must curve more strongly and consequently launch more slowly.
Finally, as we decrease $\chi_{\rm b}$, such that the streamlines are flatter to the base, the flow is already more radial and does not have to turn so quickly, hence can launch faster.
Alternatively, we consider that since the streamlines are more closely aligned to the pressure gradients, the component of the pressure gradient acting on the streamline to curve it is less.

For most regions of parameter space, including those most appropriate to describing photoevaporative winds, it is apparent that the elevation of the wind base is likely to be the most important parameter in setting the streamline curvature and the launch velocity. Only in cases of very extreme elevation, more than are considered here, would the launch velocity be suppressed to $< 0.1 c_S$.

Note that here we do not provide Mach numbers for $b\geq2$, as unlike \citet{Clarke_2016} we find that there is no possible self-similar solution for these values\footnote{\citet{Clarke_2016} were able to report a value since when integrating a solution out to a finite distance, there is always some velocity for which the singularity lies beyond that point. However, for $b\geq2$, the maximum velocity does not converge as we extend the domain of our integration.}.
Physically this represents the fact that if the density drops off as fast or faster than $r^{-2}$, then the wind must be diverging faster than spherical to ensure mass conservation, and thus would have to flow \textit{into} the launch plane. Whereas, if the density drops off slower than $r^{-2}$, then the wind must converge relative to the spherical case and can flow \textit{out} of the launch plane as required.
We provide a more mathematical discussion in Section \ref{sec:domainfilling}.

\subsection{Radial Temperature Profiles}
\label{sec:ss_nonisoT}
Having shown how self-similar streamlines may be generated for non-isothermal cases in a more general geometry above, we investigate two representative cases of radial power laws, where the temperature depends on either the spherical or cylindrical radius respectively as
\begin{align}
    T &\propto r^{-\tau} \label{eq:T_sph}\\
    T &\propto R^{-\tau} \propto \cos(\phi)^{-\tau} r^{-\tau} \label{eq:T_cyl}.
\end{align}
We provide explicit versions of equations \ref{eq:odes_0} to \ref{eq:odes_g2} for these two cases in Appendix \ref{sec:a2} and move straight to summarising our results.

\subsubsection{Results}
The resulting maximum Mach numbers are listed as the analytic values in Table \ref{tab:Mb_niT}. For comparison, we include the value for the isothermal case with the same $b$ and $\phi_{\rm b}$. In each case the effect of the temperature gradient is to lower the Mach numbers by no more than $10$ per cent when $\tau=0.25$, and up to $10-20$ per cent for $\tau=0.5$.
A decrease is to be expected - the temperature gradient at the wind base increases the outward pressure force, which makes the radius of curvature smaller and the velocities lower. Correspondingly, this effect is stronger the greater the value of $\tau$; additionally, we see stronger decreases when temperature scales with cylindrical radius.

\begin{figure}
    \centering
    \includegraphics[width=0.8\linewidth]{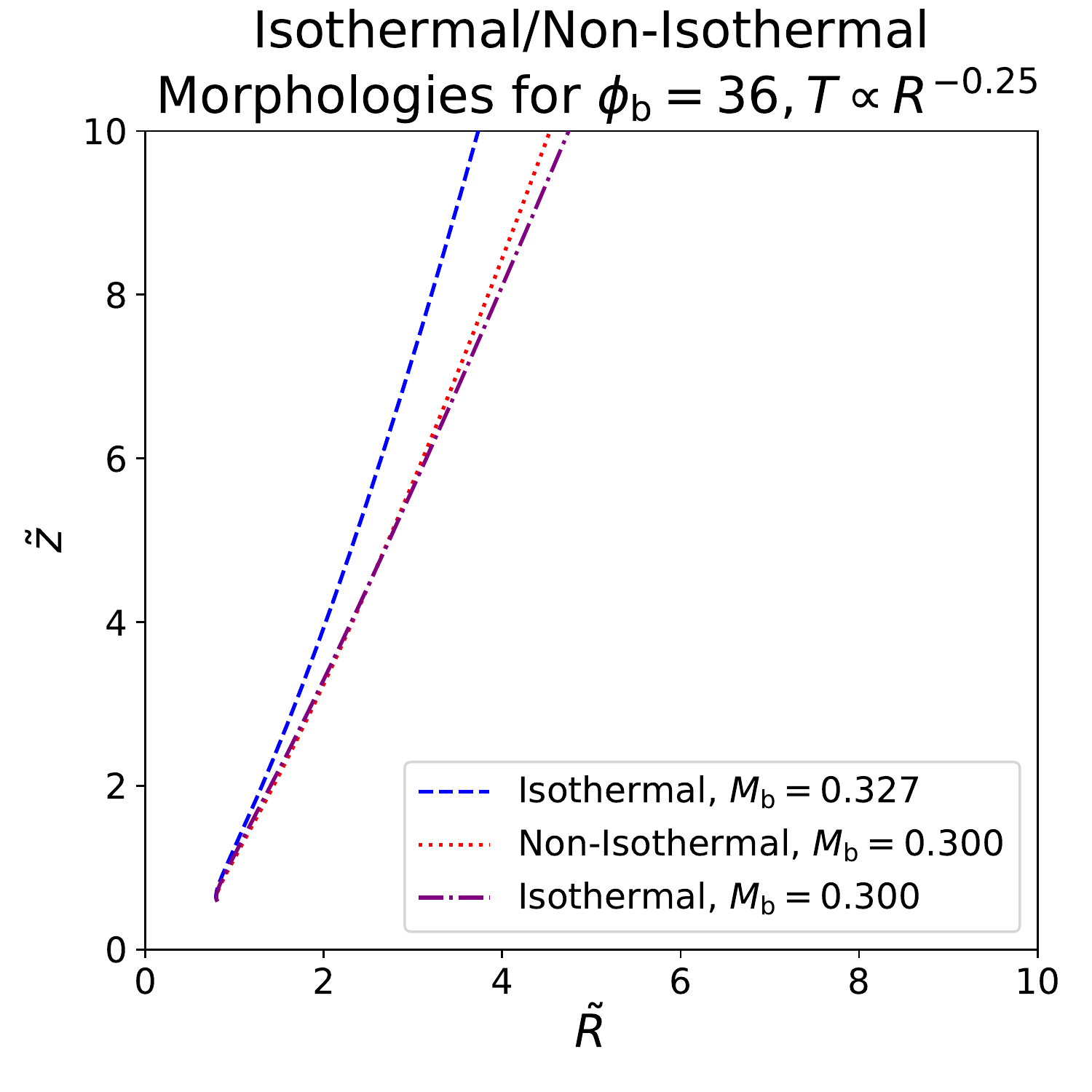}
    \caption{Comparison of the isothermal and non-isothermal ($\tau=0.25$) analytic streamline for $\phi_{\rm b}$ showing the relatively weak effects of temperature gradients. The blue dashed line is the isothermal streamline for the isothermal $\mathcal{M}_{\rm b, max}=0.327$, while the red dotted line indicates the non-isothermal streamline at its appropriate $\mathcal{M}_{\rm b, max}=0.300$. We also show, as the purple dot-dashed line, the isothermal streamline $\mathcal{M}_{\rm b}=0.300$.}
    \label{fig:streamline_comparison_analytic}
\end{figure}

To determine the effect of non-isothermality on the streamline morphologies we plot as an example a comparison of different streamlines in the $\phi_{\rm b}=36^\circ$ case in Figure \ref{fig:streamline_comparison_analytic}.
We show the appropriate streamline in the cylindrical case with $\tau=0.25$ and its maximum Mach number $0.300$.
We also include the streamline for $b=1.5$ in the isothermal case, with its appropriate Mach number $0.327$.
Finally, in order to isolate the effects of the temperature gradients from the resultant slower launch, we show an isothermal streamline with the Mach number reduced to match the non-isothermal case.

It can be seen that \textit{both} of the slower streamlines curve more strongly at the base, resulting in the streamlines passing through a lower $z$ at a given $R$.
Indeed, the effect of non-isothermality is almost completely explained by the reduction in Mach number: the isothermal case does not seriously over-estimate the radius of curvature at the base, and further deviations only set in at rather large radii when the non-isothermal solution curls upwards relative to the isothermal streamline in order to fill the spatial domain (c.f. Section \ref{sec:domainfilling}).
Note that the streamlines are also more vertical for the elevated base than for to midplane launches \citep[see][]{Clarke_2016}.

Overall, we conclude that for realistic temperature variations, the impact on the launch Mach numbers and morphologies of the flow is rather small, and hence the validity of our results should be relatively insensitive to the heating and cooling uncertainties and thus the finer picture of the thermal structure of the wind (save for any role these processes play in setting the elevation of the base).

\begin{table}
    \centering
    \caption{The mass weighted average launch Mach numbers measured for a range of non-isothermal models.}
    \label{tab:Mb_niT}
    \begin{tabular}{c|c|c|c|c|c}
    \hline
    & & \multicolumn{4}{c}{$\mathcal{M}_{\rm b}$} \\
    $\phi_{\rm b}$     &    Temperature    &   Analytic &   Isothermal  &   $u$ constant    &   $\mathcal{M}$ constant \\
    \hline
    0     & $r^{-0.25}$ &   0.522   &   0.555    &   0.528   &   0.522\\
    0     & $R^{-0.25}$ &   0.506   &   0.555    &   0.512   &   0.507\\
    36    & $r^{-0.25}$ &   0.322   &   0.327    &   0.330   &   0.329\\
    36    & $R^{-0.25}$ &   0.300   &   0.327    &   0.309   &   0.308\\
    0     & $r^{-0.5}$  &   0.472   &   0.555     &   0.495   &   - \\
    0     & $R^{-0.5}$  &   0.449   &   0.555     &   0.467   &   - \\
    36    & $r^{-0.5}$  &   0.305   &   0.327     &   0.318   &   - \\
    36    & $R^{-0.5}$  &   0.270   &   0.327     &   0.282   &   - \\
    \hline
    \end{tabular}
\end{table}

\section{Hydrodynamic Simulations of Scale Free Winds}
\label{sec:ss_hydro}
Having established new predictions for the launch velocities of self-similar winds for more general base geometries and non-isothermal temperature profiles, we wish to benchmark these against hydrodynamic simulations. We therefore present \textsc{FARGO3D} \citep{FARGO_reference} simulations in order to assess whether the result that scale-free winds adopt the solution with $\mathcal{M}_{\rm b,max}$ applies for generic values of the angles $\phi_{\rm b}$ and $\chi_{\rm b}$ and for temperature gradients $\tau>0$.

\subsection{Description of FARGO3D Setup}
Since for a direct comparison, we desire a scale-free scenario, our setup uses no gravitational forces, with the azimuthal velocity set to $0$ to eliminate centrifugal forces arising from orbital motion.

We used a 2D spherical grid with $220$ cells logarithmically spaced in $r$ between $r=0.01$ and $r=10$ and $50\left( 1 - \phi_{\rm b} / 90^\circ \right)$ cells spaced linearly between $\theta=0$ and $\theta=90-\phi_{\rm b}$ (where $\theta$ is now the usual colatitudinal angle), such that the grid cells are approximately square and we have a fixed constant angular resolution.
While this is lower than the resolution of \citet{Clarke_2016}, we tested that this did not affect our results.
Note that since these simulations are scale free then here the radius values quoted have no particular meaning but are simply relative. 

The launch plane at $\theta=90-\phi_{\rm b}$ was treated by having constant perpendicular velocity across the boundary, with the parallel component set to 0 and the density set to $\rho = r^{-b}$.
When $\chi_{\rm b} \neq \pi/2$, the perpendicular velocity $u_\theta$ at the launch plane is still imposed to be constant across the boundary. We then use its value to set the parallel component $u_r=-u_\theta \cot(\chi_{\rm b})$ (the sign accounts for the fact that the positive $\theta$ direction is directed into the plane; we are interested in winds where $u_\theta<0$, $u_r>0$.
At the polar axis, we used a reflecting boundary.

For both radial boundaries, we required that the components of velocity should be constant across the boundary but that the density follow the same imposed power law slope as at the base.
We argue that this is the correct boundary condition to use if we wish to seek perfect agreement with the self-similar solution since, as mentioned in Section \ref{sec:mathematical_extension}, at a fixed angle we probe equivalent points on adjacent streamlines and so the density should simply scale with the density at their bases.
Likewise the velocity profile is constant at the base and thus should be at all angles. We checked that our setup recovered the perpendicular, `flat plane' ($\phi_{\rm b}=0$), cases, and found that these boundary conditions have the effect of reducing the deviations from self-similarity near the boundaries \citep[c.f.][Figure 4]{Clarke_2016}. This is also apparent in Figures \ref{fig:ss_geometry_Mb}, where the Mach numbers remain exactly flat for all $r$.

\textsc{FARGO3D} allows two equations of state: an adiabatic case (in which it evolves the energy), and a locally isothermal one. We use the latter, in which \textsc{FARGO3D} stores the sound speed (such that the pressure is calculated as $P = c_S^2 \rho$), which it does not evolve in time; its value is fixed as part of the initial conditions. For our globally isothermal cases (i.e. $\tau=0)$, we set $c_S=1$ everywhere, while for our power law temperature profiles, we follow equation \ref{eq:cs_general}.

\subsection{Generalised Geometry}
\begin{figure*}
    \centering
    \includegraphics[width=\linewidth]{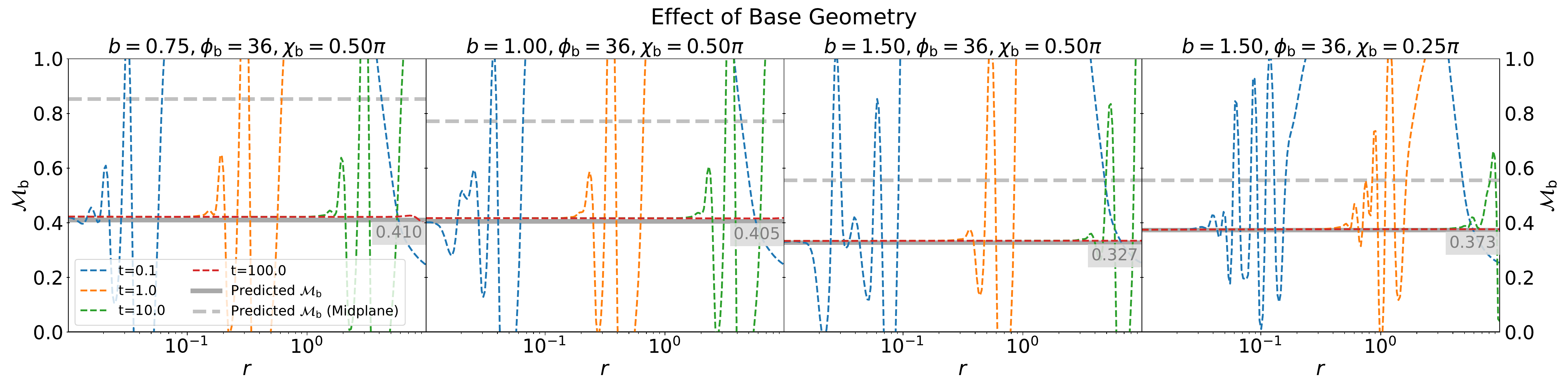}
    \caption{Comparison of the launch Mach numbers $\mathcal{M}_{\rm b}$ for scale-free wind models in our generalised geometry. From left to right: winds launched perpendicularly from elevated bases ($\phi_{\rm b}=36^\circ$) with density power law slopes $b=0.75,1.00,1.50$, and a wind with $b=1.5$ launched at $\chi_{\rm b}=0.25\pi$ from a base elevated by $\phi_{\rm b}=36^\circ$.
    The coloured dashed lines indicate output hydrodynamic simulations at various times. 
    The darker grey band represents the predicted $\mathcal{M}_{\rm b, max}$ from the self-similar models and the grey label shows its value. To emphasise the reduction that the elevation of the base causes, we also illustrate the $\mathcal{M}_{\rm b, max}$ for winds of the same $b$ launched perpendicularly from the midplane as the lighter grey, dashed, band.}
    \label{fig:ss_geometry_Mb}
\end{figure*}

\begin{figure*}
    \centering
    \includegraphics[width=\linewidth]{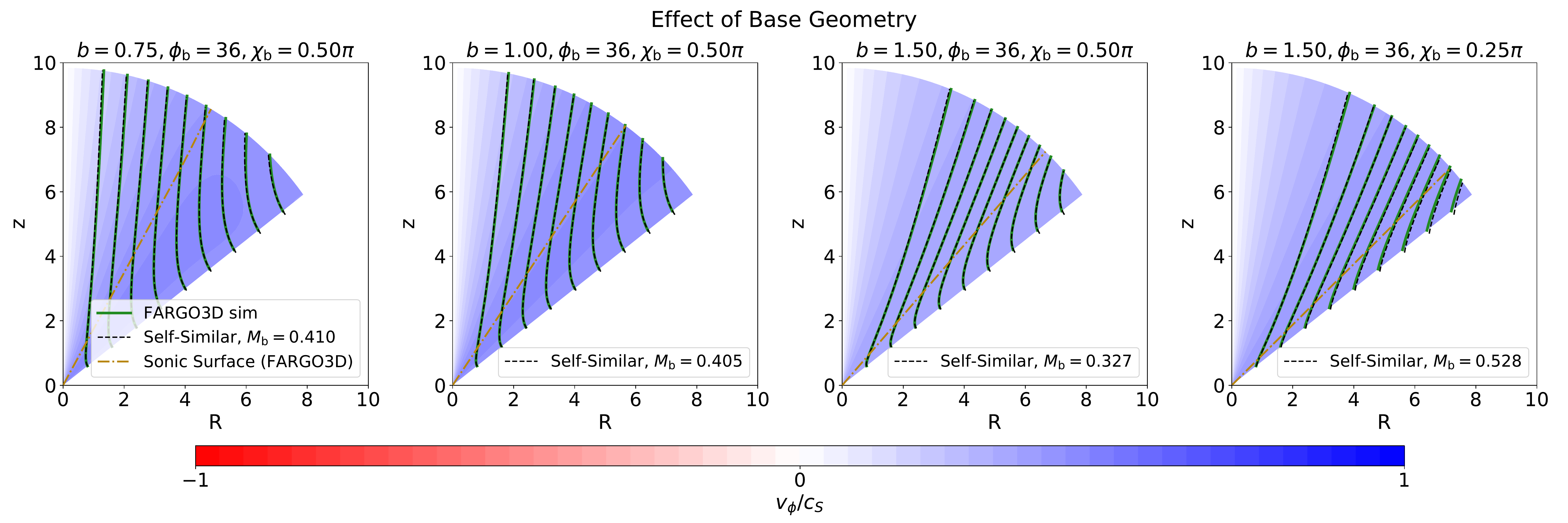}
    \caption{Comparison of the streamline morphology for wind models in our generalised geometry demonstrating complete agreement of scale-free hydrodynamic simulations with self-similar solutions. From left to right: winds launched perpendicularly from elevated bases ($\phi_{\rm b}=36^\circ$) with density power law slopes $b=0.75,1.00,1.50$, and a wind with $b=1.5$ launched at $\chi_{\rm b}=0.25\pi$ from a base elevated by $\phi_{\rm b}=36^\circ$.
    The solid green lines are the streamlines retrieved from the final outputs of the hydrodynamic models, whereas the black dashed lines are those found by numerical integration of equations \ref{eq:odes_0}-\ref{eq:odes_g2} for $\tau=0$, $\mathcal{C}=1$. The gold dashed-dotted line shows the sonic surface. The background is coloured according to the velocity in the $\phi$ direction.}
    \label{fig:ss_geometry_streamlines}
\end{figure*}

In Figures \ref{fig:ss_geometry_Mb} and \ref{fig:ss_geometry_streamlines}, we examine the launch Mach number $\mathcal{M}_{\rm b}$ and streamlines respectively of winds launched from a base elevated by $\phi_{\rm b}=36^\circ$ to the midplane.
We again choose this illustrative value to match photoevaporation models \citep[e.g.][]{Picogna_2019,Wang_2017} since the qualitative results were found not to vary with elevation.

Figure \ref{fig:ss_geometry_Mb} shows the $\mathcal{M}_{\rm b}$ of a subset of our simulations as a function of radius at a range of times.
We focus on the three cases of $b$ examined in \citet{Clarke_2016}: $0.75,1.00,1.50$ which are shown from left to right.
For $b=1.5$ we also show in the right-most panel a model that is not launched perpendicularly; we choose an illustrative value of $\chi_{\rm b}=0.25\pi$, motivated by the approximate extreme value shown by the innermost streamlines in the simulations of \citet{Picogna_2019} and \citet{Wang_2017}.

It is apparent that for all density profiles, the winds converge to a constant $\mathcal{M}_{\rm b}$ value (represented by the dark grey band) from the inside out on a timescale approximately proportional to $r$.
Moving to an elevated base has not degraded the result that the hydrodynamic simulations converge to a value of $\mathcal{M}_{\rm b}$ that matches the $\mathcal{M}_{\rm b, max}$ predicted from self-similar models (which is typically much lower than that of the winds launched from the midplane: light grey dashed band).
Any small differences can be attributed to the fact that the simulation outputs plotted are derived at the centre of the grid cell closest to the base, when the base itself is at the imposed angle, which we confirmed by increasing the resolution and seeing that the $\mathcal{M}_{\rm b}$ indeed converges more closely towards the predicted value.

Moreover this agreement holds for the non-perpendicularly launched wind in the right-most panel of Figure \ref{fig:ss_geometry_Mb}. Note that for a direct comparison (and as would be relevant to interpreting mass loss rates), we plot only the component of velocity perpendicular to the plane. Therefore though the Mach number is increased mildly above the $\mathcal{M}_{\rm b, max}$ for the perpendicular case, the enhancement is smaller than would be immediately inferred from Figure \ref{fig:Mb_contours}, which shows the total velocity.
We confirmed that the agreement shown here between the predicted $\mathcal{M}_{\rm b,max}$ and the  $\mathcal{M}_{\rm b}$ found in the hydrodynamic simulations held for a range of $\chi_{\rm b}$, both for winds launched from the midplane and the elevated base.

In Figure \ref{fig:ss_geometry_streamlines} we show a comparison of the streamlines integrated from the hydrodynamic simulations shown in Figure \ref{fig:ss_geometry_Mb} to those obtained in Section \ref{sec:numerical_solutions} (scaled by the base radius), for the appropriate $M_{\rm b,max}$. The agreement is excellent for all streamlines showing that the self-similarity is adopted throughout the domain. Moreover, the sonic surface is a surface of constant $\phi$. It is clear from these models that as described in the previous section, the higher the value of $b$, the more rapidly the streamlines curve over. 

The conclusions of \citet{Clarke_2016} for scale free winds thus also apply to elevated bases and non-perpendicular launches: self-similar solutions with $\mathcal{M}_{\rm b, max}$ are generally applicable for \textit{any} scale-free isothermal wind, regardless of base geometry.

\subsection{Radial Temperature Profiles}
Having derived and computed self-similar solutions in Section \ref{sec:ss_nonisoT} for temperature profiles that are power laws in either spherical radius (equation \ref{eq:T_sph}) or cylindrical radius (equation \ref{eq:T_cyl}), we also benchmark the behaviour of equivalent hydrodynamic simulations.

\begin{figure*}
    \centering
    \includegraphics[width=\linewidth]{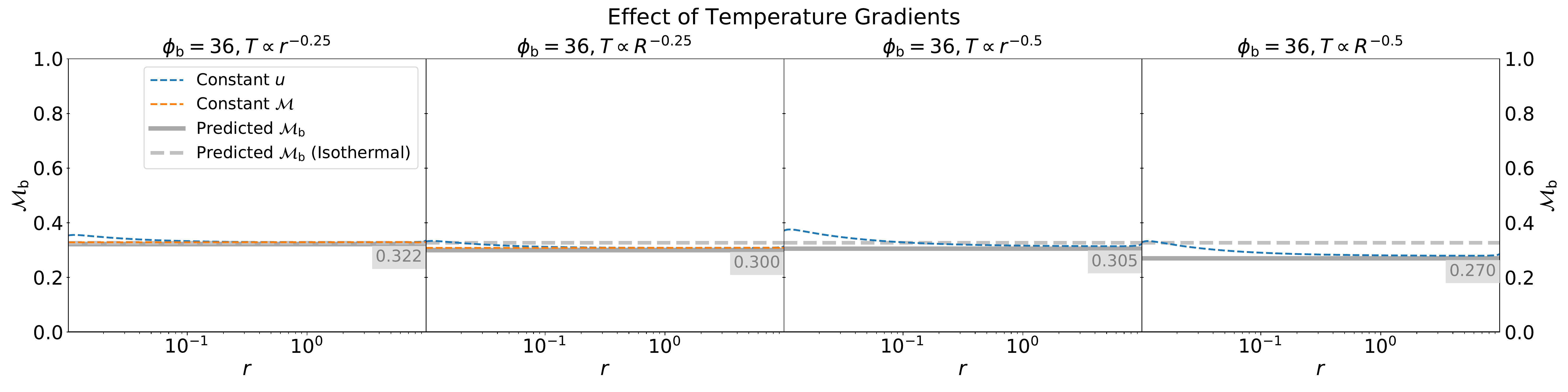}
    \caption{Comparison of the launch Mach numbers $\mathcal{M}_{\rm b}$ (measured with respect to the local sound speed) for non-isothermal winds with density power law index $b=1.5$ and temperature power law index $\tau=0.25$ (first and second panels, spherical and cylindrical respectively) and $\tau=0.5$ (third and fourth panels, spherical and cylindrical respectively) for discs with wind bases elevated by $\phi_{\rm b}=36^\circ$. The blue lines show simulations with constant velocity imposed at the radial boundaries; the orange lines use a constant Mach number. The darker grey band represents the predicted $\mathcal{M}_{\rm b, max}$ from the self-similar models and the grey label shows its value, while the lighter grey, dashed, band shows the $\mathcal{M}_{\rm b, max}$ for isothermal winds with otherwise equivalent parameters.}
    \label{fig:niT_Mb}
\end{figure*}

In Figure \ref{fig:niT_Mb}, we observe that the Mach number - measured with respect to the local sound speed - achieved in the simulations (as represented by the blue dashed line) is indeed nearly constant in radius - for both the spherical and cylindrical power laws and for both values of $\tau$ - as we have argued is appropriate in the self-similar case.
In the outer disc at $\tilde{r} \gtrsim 1$, the Mach number appears to correspond well with the values predicted from self-similar models as listed in Table \ref{tab:Mb_niT}.
$\mathcal{M}_{\rm b}$ does however, seem to increase as we move towards the inner disc. This is an artefact of the boundary conditions, since these are designed to impose constant \textit{velocity} components across the radial boundaries of the simulation grid.
Therefore we reran the models with $\tau=0.25$\footnote{With these boundaries, $\tau=0.5$ did not achieve a steady solution.}
with a boundary condition designed to impose constant Mach numbers across the radial boundaries.
The results are presented as the orange lines in Figure \ref{fig:niT_Mb} and indeed eliminate the deviation from constant $\mathcal{M}_{\rm b}$ in the inner disc.

We see that the coloured lines from the hydrodynamic simulations are very close to the predictions of the corresponding non-isothermal analytic solutions (dark grey lines), which involve slightly smaller launch velocities than in the isothermal case (pale grey lines).
To quantify the agreement further, we measure the mass-weighted average $\mathcal{M}_{\rm b}$ from each simulation and report the values for both sets of boundary conditions in Table \ref{tab:Mb_niT}; we also include values for winds with $\phi_{\rm b}=0$.
The Mach numbers are generally within $0.01-0.02$ of the predicted values, and thus, as expected, typically decrease by on the order of 10 per cent from the isothermal values; any small discrepancies are due to boundary effects. 

\begin{figure*}
    \centering
    \includegraphics[width=\linewidth]{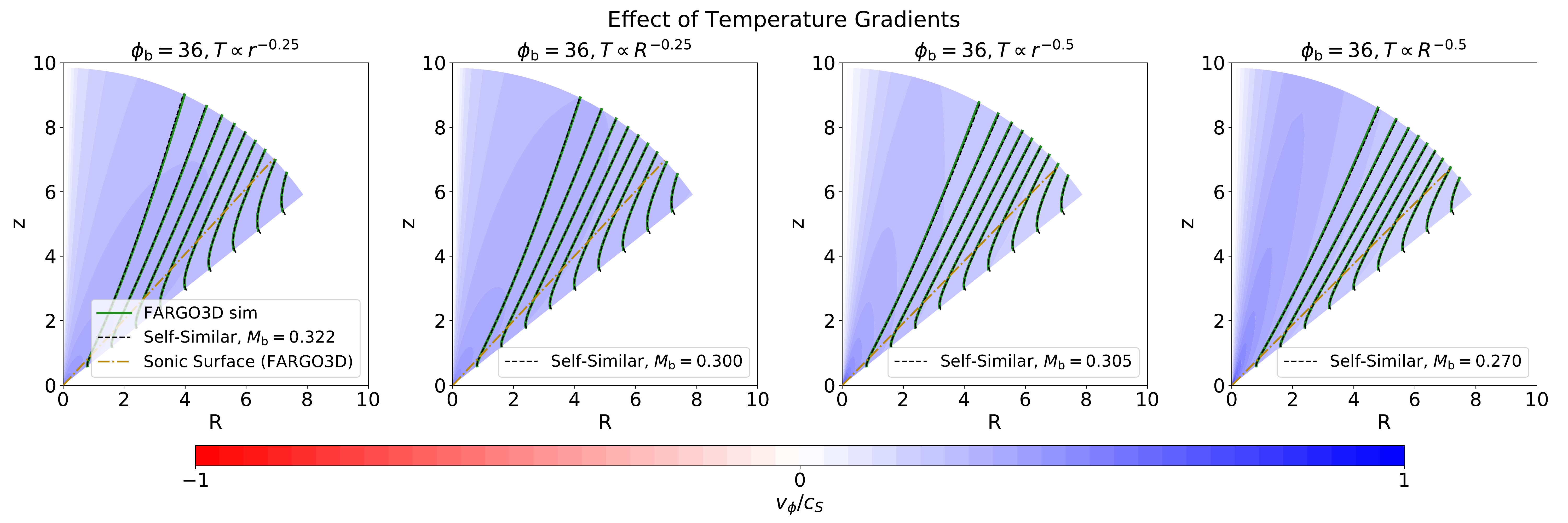}
    \caption{As with the $b=1.5$ case in Figure \ref{fig:ss_geometry_streamlines} but for the models in Figure \ref{fig:niT_Mb} which include additional temperature gradients. The black dashed lines are the analytical streamlines for the corresponding Mach numbers reported in Table \ref{tab:Mb_niT}.}
    \label{fig:niT_streamlines}
\end{figure*}

We also check the morphology of the streamlines for our non-isothermal models, by plotting the $\tau=0.25$ simulations with the constant Mach number boundaries and the $\tau=0.5$ simulations in Figure \ref{fig:niT_streamlines}. This shows that for both power laws and gradients we also have excellent agreement between the hydrodynamic simulations and our model predictions.

\section{Domain Filling}
\label{sec:domainfilling}

Having demonstrated that no matter the location of the wind base, angle of launch, or temperature structure, the hydrodynamic solution matches the self-similar solution with $\mathcal{M}_{\rm b} = \mathcal{M}_{\rm b,max}$, we now consider why this must be the case. As we shall demonstrate, this solution corresponds to the one for which the streamlines asymptote to the maximum possible angle, $\pi/2$, with the midplane; consequently for $\mathcal{M}_{\rm b} < \mathcal{M}_{\rm b,max}$, the winds cannot fill the computational domain.

As $\tilde{r}\to\infty$, the angle with the midplane, $\phi$, which is bound to lie between $0$ and $\pi/2$, cannot change indefinitely.
Hence, at large radii, we expect the streamlines to become asymptotically straight lines of constant $\phi=\phi_\infty$.
Since the streamlines become straight, then note that in this regime the radius of curvature $|\tilde{R}_{\rm eff}| \to \infty$.

\begin{figure}
    \centering
    \includegraphics[width=\linewidth]{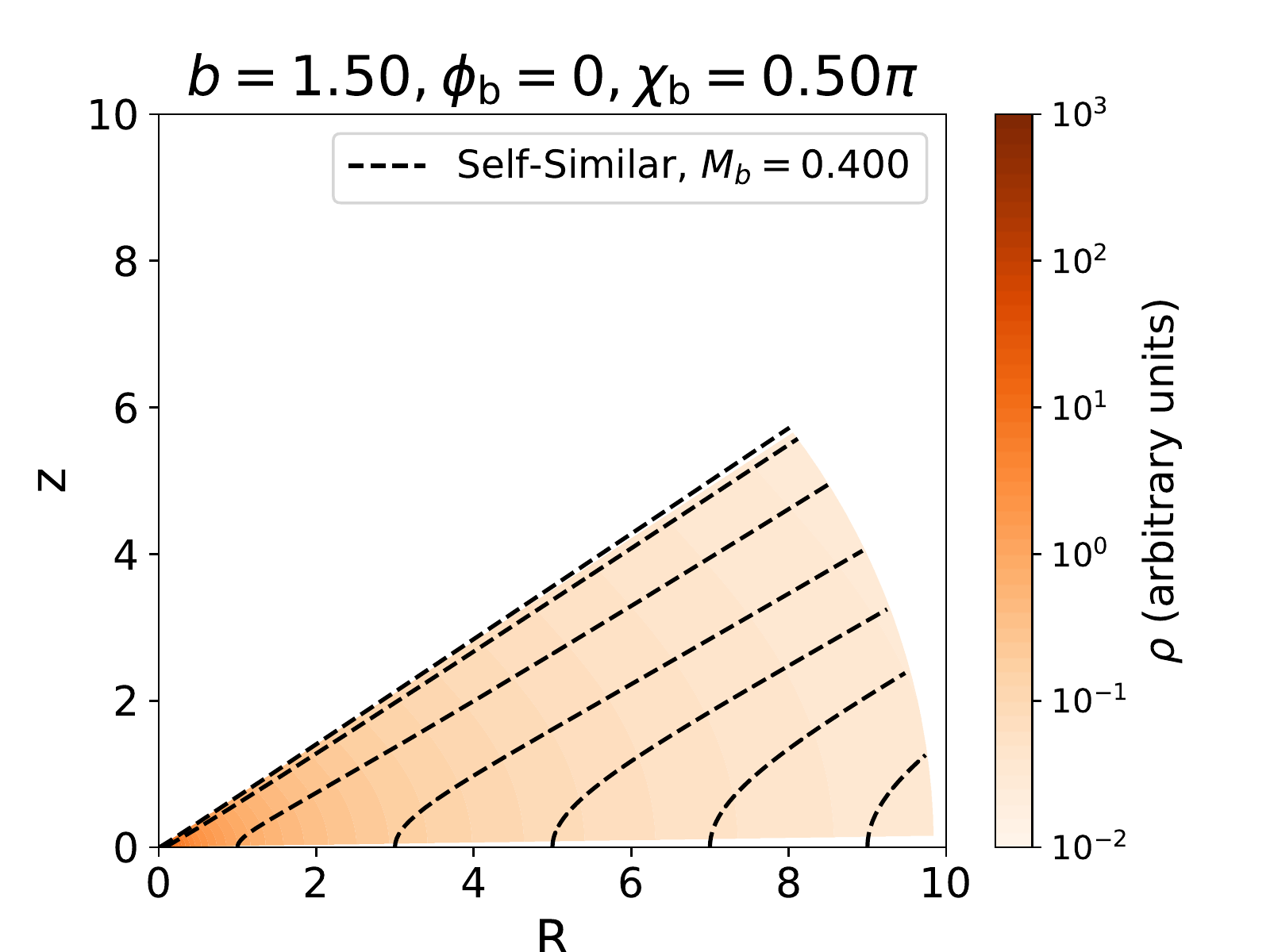}
    \caption{The density contours inferred from a self-similar wind model which launches more slowly than the maximum Mach number, with various streamlines overplotted as the black dashed lines. The streamlines asymptote to an angle $\approx 36^\circ$ with the midplane, leading to a sharp discontinuity in the density at this angle.}
    \label{fig:density_nonfilling}
\end{figure}

Since the domain of the problem - and hence both our FARGO3D simulations and those carried out by \citet{Clarke_2016} - extends all of the way to the z-axis at $\phi=\pi/2$, this means that were all the streamlines to asymptote to $\phi_\infty<\pi/2$, the wind could not fill the domain.
The regions at $\phi>\phi_\infty$ would thus end up empty and provide no resistance to being filled by the wind.
In the limit that $|\tilde{R}_{\rm eff}| \to \infty$ as $\phi \to \phi_\infty$, the pressure gradient perpendicular to the streamlines (right-hand side of equation \ref{eq:eom} should be $0$ in the self-similar model.
However the empty region thus creates a discontinuity in the density; in other words this creates strong perpendicular pressure (density) gradients at $\phi_\infty$ that would curve the streamlines upwards into the empty region.
We thus expect $\phi_\infty<\pi/2$ to be unstable.
We provide an illustrative example for a model of a self-similar wind launched from the midplane with $\mathcal{M}_{\rm b} = 0.4 < \mathcal{M}_{\rm b, max}$ in Figure \ref{fig:density_nonfilling}, for which the wind only fills $\phi \lesssim 36^\circ$.
We note that instead if the wind extends up to $\phi=\pi/2$, there must by default be no pressure gradient in the angular direction due to the symmetry about the z-axis, which is consistent with the self-similar picture. Thus we conclude that self-similar streamlines with $\phi_\infty<\pi/2$ would not be  self-consistent.

Therefore, we wish to know what $\phi_\infty$ any particular streamline with $\mathcal{M}_{\rm b} \leq \mathcal{M}_{\rm b, max}$ tends to as $\tilde{r}\to\infty$.
However, since the integration we performed was over a finite range (up to $\tilde{y}=1000$), we need a way to extrapolate these streamlines to infinity from the largest $\tilde{r}$ and $\phi$ achieved.

\subsection{Limiting Behaviour of Streamline Morphology}
We conduct this analysis for the simpler case of an isothermal wind, and start by considering the limiting behaviour of the streamline morphology. Since $\chi$ measures the angle between the streamline tangent at a given point and the radial direction, if $\chi\neq0$ then $\phi$ is changing; the relation between the two is:
\begin{equation}
    {\rm d}\tilde{r} \tan(\chi) = \tilde{r}{\rm d}\phi.
    \label{eq:chi_dphi}
\end{equation}
Hence for asymptotically straight streamlines, we must have $\sin(\chi) \approx \tan(\chi) = \frac{{\rm d} \phi}{{\rm d} \ln\tilde{r}} \to 0$ as $\tilde{r}\to\infty$.

As described above, in the radial limit, the pressure gradient perpendicular to the streamlines must also be 0. From equation \ref{eq:eom}, this is the condition that
\begin{equation}
        \frac{b}{\tilde{r} \sin(\chi)} - \mathcal{M}_{\rm b}^2 \tilde{u} \frac{{\rm d}\tilde{u}}{{\rm d}\tilde{z}} \cot(\chi) \sin(\theta) = 0.
        \label{eq:eom_Reffinf}
\end{equation}
Since as $\chi\to0$ we may assume that $\cos(\chi)\approx1$ then ${\rm d}\tilde{z} = {\rm d}\tilde{s}\sin(\theta) \approx {\rm d}r\sin(\theta)$, and this rearranges to give
\begin{equation}
         \frac{{\rm d} \tilde{u}^2}{{\rm d} \ln(\tilde{r})}
        = 2\frac{b}{\mathcal{M}_{\rm b}^2},
        \label{eq:du2}
\end{equation}
which integrates to give the asymptotic form of the velocity
\begin{equation}
    \tilde{u}^2 \sim \frac{2b}{\mathcal{M}_{\rm b}^2} \ln(\tilde{r}).
    \label{eq:asymptotic_u}
\end{equation}

In the isothermal case, when equation \ref{eq:Bernoulli_general_P} integrates to give to the constancy of the Bernoulli function, \cite{Clarke_2016} showed that the density could be expressed as
\begin{equation}
    \tilde{\rho} = \exp\left(-\frac{\mathcal{M}_{\rm b}^2}{2}(\tilde{u}^2-1)\right).
    \label{eq:Bernoulli_iso}
\end{equation}
Substituting equation \ref{eq:asymptotic_u} in equation \ref{eq:Bernoulli_iso}, the asymptotic behaviour of the density becomes $\rho \propto r^{-b}$, in agreement with our statement in section \ref{sec:mathematical_extension} about the variation at fixed $\phi$:
\begin{equation}
    \tilde{\rho} \sim \exp\left(\frac{\mathcal{M}_{\rm b}^2}{2}\right) \tilde{r}^{-b}.
\end{equation}
Using equation \ref{eq:masscons_mdot} (in its self-similar form),
we then proceed to write the angular terms as
\begin{equation}
    \sin(\chi) \cos(\phi) \sim \sin(\chi_{\rm b}) \cos(\phi_{\rm b}) \frac{1}{\sqrt{2b}} \mathcal{M}_{\rm b} \exp\left( -\frac{\mathcal{M}_{\rm b}^2}{2} \right) \frac{\tilde{r}^{b-2}}{\sqrt{\ln(\tilde{r})}}.
    \label{eq:asymptotic_sinchi}
\end{equation}
Note that since $\cos(\phi)$ is bounded, to achieve $\sin(\chi)\to0$, it is necessary to have $b<2$ such that the right hand side is a decreasing function of $\tilde{r}$ at large radii\footnote{Note however that this is not a sufficient condition for $\sin(\chi)\to0$ yet since if $\phi\to\pi/2$, then $\cos(\phi)\to0$ also.}. This demonstrates why we cannot have a self-similar solution for $b \geq 2$: for such values, the streamlines cannot tend to be straight but must increase in curvature\footnote{Strictly speaking, for $b=2$ here we do have a decreasing function of $\tilde{r}$, but the small correction to the velocity gradients due to curvature forces destabilises this case too.}, which reinforces the physical argument made in section \ref{sec:numerical_solutions} that they have to diverge in order to conserve mass. 

Approximating $\tan(\chi)\approx\sin(\chi)$, we now combine equations \ref{eq:chi_dphi} and \ref{eq:asymptotic_sinchi} and integrate (by substitution $t=\sqrt{(2-b)\ln(\tilde{r})}$) using the boundary condition $\phi\to\phi_\infty$ as $\tilde{r}\to\infty$. This gives us the polar equation of our streamlines at large radius:
\begin{align}
    \sin(\phi) &\sim \sin(\phi_\infty) - A \sin(\chi_{\rm b}) \cos(\phi_{\rm b})  \erfc\left(\sqrt{(2-b)\ln(\tilde{r})}\right)
    \label{eq:sin_erfc}, \\
    A(b,\mathcal{M}_{\rm b}) &= \sqrt{\frac{\pi}{2b(2-b)}} \mathcal{M}_{\rm b} \exp\left( -\frac{\mathcal{M}_{\rm b}^2}{2} \right).
\end{align}

\subsection{Morphology of Slow Winds}
\begin{figure}
    \centering
    \includegraphics[width=\linewidth]{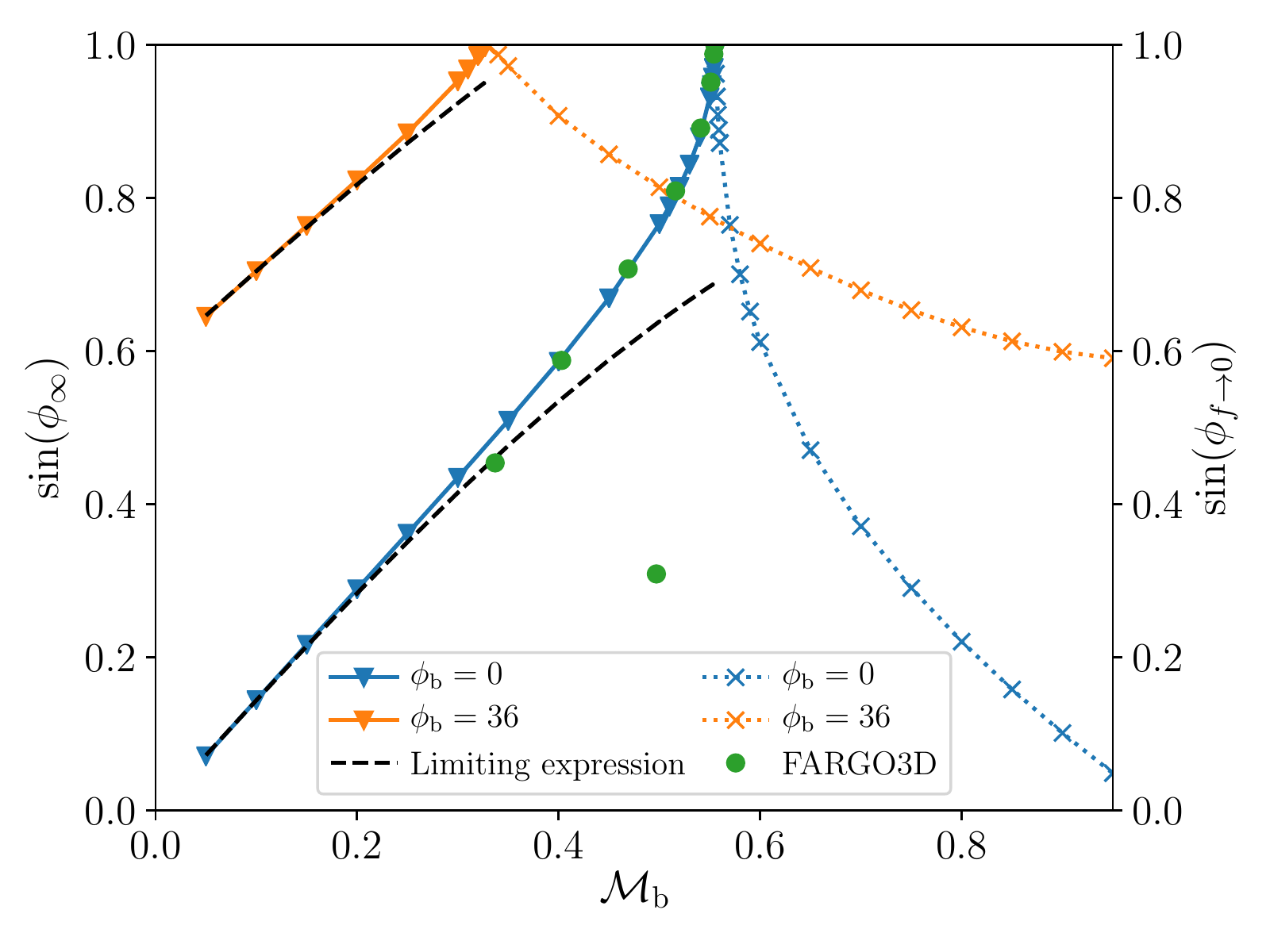}
    \caption{The key angles describing the behaviour of self-similar wind solutions at large radius. The left-hand axis (solid lines, triangles) measures the angle to which the streamlines asymptote $\phi_\infty$ for self-similar midplane winds (blue), self-similar winds with $\phi_{\rm b}=36^\circ$ (orange) and winds in hydrodynamic simulations on a restricted domain (green circles) as a function of $\mathcal{M}_{\rm b}\leq\mathcal{M}_{\rm b, max}$. The right-hand axis (dotted lines, crosses) plots the angle $\phi_{f\to0}$ at which self-similar solutions with $\mathcal{M}_{\rm b}>\mathcal{M}_{\rm b, max}$ encounter a singularity. 
    The black dashed line represents a limiting expression (Equation \ref{eq:sin_erfc_lim}) which approximates $\phi_\infty$ for winds with low $\mathcal{M}_{\rm b}$ (and thus works particularly well for $\phi_{\rm b}=36^\circ$).}
    \label{fig:phiinfty_Mb}
\end{figure}

Now, as desired, for any streamline calculated for a given value of $\mathcal{M}_{\rm b} \leq \mathcal{M}_{\rm b, max}$, we use equation \ref{eq:sin_erfc} to estimate $\sin(\phi_\infty)$ from the maximum $\tilde{r}$ and $\phi$ reached in our integration.
The results are presented in Figure \ref{fig:phiinfty_Mb} for isothermal cases with $b=1.5$, $\chi_{\rm b}=\pi/2$ and $\phi_{\rm b}=0,36^\circ$ in blue and orange respectively. 
$\sin(\phi_\infty)$ plotted on the left-hand axis using triangle markers and solid lines.

We see clearly that for both values of $\phi_{\rm b}$, as $\mathcal{M}_{\rm b}$ is reduced from its maximal permitted value, the $\phi_\infty$ reached by the streamlines decreases from $\pi/2$, with $\phi_\infty=\pi/2$ if and only if $\mathcal{M}_{\rm b} = \mathcal{M}_{\rm b,max}$.
That is to say that the slower the wind the lower (in $\phi$) the surface to which it asymptotes.
Physically this happens since when the domain filled by the wind is restricted, the winds must curve more strongly as they have less room in which to become radial, just as in the case of an elevated base.
Therefore, the reason that $\mathcal{M}_{\rm b, max}$ is the preferred launch Mach numbers is because it is the only solution which fills the full domain.

Figure \ref{fig:phiinfty_Mb} also includes black dashed lines representing a limiting expression found by evaluating Equation \ref{eq:sin_erfc} at the base $\tilde{r}=1$, $\phi=\phi_{\rm b}$ and assuming $\chi_{\rm b}=\pi/2$,  for each $\phi_{\rm b}$ being considered:
\begin{equation}
    \sin(\phi_\infty) = \sin(\phi_{\rm b}) + \cos(\phi_{\rm b}) \sqrt{\frac{\pi}{2b(2-b)}}  \mathcal{M}_{\rm b} \exp\left( -\frac{\mathcal{M}_{\rm b}^2}{2} \right)
    \label{eq:sin_erfc_lim}
\end{equation}
This is useful as it provides a good estimate of the relationship between $\mathcal{M}_{\rm b}$ and $\phi_\infty$ in the case that the domain filled by the wind is restricted, either because of an elevated base $\phi_{\rm b}=36$ or a slow launch. This comes about because these initially more highly curved solutions become radial more quickly, while still near the base.

\subsection{Relationship with Invalid Solutions}
We can also use this framework -
considering the rate of change of $\phi$, expressed in terms of $\sin(\chi)\approx\tan(\chi)$ - 
to understand why the Mach number above which there are no valid solutions coincides with that for which the solutions fill the domain.
At a given (large) radius and $\phi$, $\sin(\chi) \propto \mathcal{M}_{\rm b} \exp\left( -\frac{\mathcal{M}_{\rm b}^2}{2} \right)$, which is an increasing function of $\mathcal{M}_{\rm b}$ in the subsonic regime.
Thus for lower Mach numbers than the domain filling solution, the rate of change of $\phi$ would be too small to reach $\phi=\pi/2$.
Conversely for any faster launch
the solution would try to curve up too steeply to reach impossibly large $\sin(\phi)>1$.

Practically, these solutions must therefore break down; instead they encounter the critical point where $f_1+f_2=0$ and the velocity gradient diverges \citep{Clarke_2016}.
The connection between $f\to0$ and high $\sin(\chi)$ can be made explicit by considering an equivalent criterion
expressed in terms of the ratio of the two terms:
\begin{equation}
    \mathcal{F} = \frac{|f_1|}{f_2} = (\mathcal{M}_b^2 \tilde{u}^2 -1) \tan^2(\chi),
\end{equation}
such that $\mathcal{F}\to 1$ represents the singularity.
Note that since $\tilde{u}$ will be monotonically increasing (though potentially very mildly) then for $\mathcal{F}$ to remain less than $1$, $\tan(\chi) = \frac{{\rm d} \phi}{{\rm d} \ln\tilde{r}}$ must be monotonically decreasing towards $0$ - again we see the validity of the solution is determined by the streamlines becoming asymptotically radial.

For comparison therefore, we also plot the value of $\phi$ where these solutions with $\mathcal{M}_{\rm b} > \mathcal{M}_{\rm b,max}$ reach the singularity on the right-hand axis of Figure \ref{fig:phiinfty_Mb} using crosses and dotted lines.
We see that indeed the faster the wind is launched, the larger $\sin(\chi)$ becomes and so the sooner the singularity is encountered \citep{Clarke_2016}.

\subsection{Hydrodynamic Simulations on a Restricted Domain}
Although with the simulations presented so far, the winds must adopt $\phi_\infty=\pi/2$, we can imagine a scenario, such as some separate magnetic or stellar outflow from smaller radii, that acts to provide the geometrical constraint on the region occupied by disc winds. For example, \citet{Hollenbach_1994, Richling_1997} consider the effects of a strong stellar wind (as appropriate for O/B type stars) on the EUV irradiation - and consequent ionisation balance and wind base density profile. In this scenario, pressure equilibrium is established between the ram pressure of the stellar wind, and the pressure of the disc atmosphere/wind \citep[e.g. at a `laminar slip' boundary][]{Hollenbach_1994}.

To simulate such a scenario here we simply move our reflecting boundary in our FARGO simulations from the z-axis to some lower $\phi$ values $\phi_{\rm max} = {0.45,0.4,0.35,0.3,0.25,0.2,0.15,0.1}\pi$, noting that this approach neglects any shear effects between the constraining region and the thermal wind; while this may not be entirely realistic, we use this to illustrate the validity of our interpretation of the slow winds using the most appropriate boundary conditions for the solutions.
We assume winds launched from the midplane i.e. $\phi_{\rm b}=0$.
Otherwise this set of simulations are the same as those in Section \ref{sec:ss_hydro}. The values of $\mathcal{M}_{\rm b}$ that resulted (calculated as a mass weighted average across the base) are shown as the green dots in Figure \ref{fig:phiinfty_Mb}.
These agree well for large enough $\phi_{\rm max}$, and apparently down to $\sin(\phi_{\rm max}) = 0.4-0.5$. However, more detailed inspection of the simulations shows that for $\phi_{\rm max}<0.25\pi$ the results progressively deviate from steady, self-similar, solutions, but display oscillations at small radii.
We were unable to bring these into agreement with the expected self-similar solutions by increasing the resolution in either spatial direction or in time. Nor does there appear to be some critical angle or $\mathcal{M}_{\rm b}$, consistent across choices of $b$ and $\phi_{\rm b}$, below which this behaviour manifests.
Since it is not clear that such strongly restricted scenarios are realistic, we don't consider or try to explain them any further.

We have thus shown that the solution with $\mathcal{M}_{\rm b, max}$ is a robust prediction for self-similar thermal winds, since it is the unique valid solution that fills the spatial domain (Figure \ref{fig:phiinfty_Mb}). It is only if the domain is reduced by somehow constricting the wind that self-similar winds are expected to launch more slowly than $\mathcal{M}_{\rm b, max}$.

\section{Hydrodynamic Simulations with Gravity and Centrifugal Force}
\label{sec:nss_hydro}
Beyond the agreement of the scale-free simulations with the self-similar model with maximal $\mathcal{M}_{\rm b}$, which we have shown to apply to more general scale-free winds, \citet{Clarke_2016} also demonstrated that this model provided a good prediction for the outer regions of discs even once gravity/centrifugal forces were included. We thus repeat the exercises from Section \ref{sec:ss_hydro} with gravitational potential included. Although we still assume axisymmetry and do not model the azimuthal direction, in order to provide the centrifugal force we set a non-zero azimuthal velocity at the base equal to the Keplerian value at that cylindrical radius inherited from the disc material
\begin{equation}
    u_{\rm azimuthal} = R^{-1/2}.
    \label{eq:u_azimuthal}
\end{equation}
Note that this is applied regardless of the elevation of the wind base since the corrections \citep{Nelson_2013} due to elevation above the midplane, which depend on the disc's density and temperature structure, are small.
Note further that we use units where $GM_*=1$ and $c_S=1$, such that the radius is now expressed in units of $r_G$. In addition, we found that we now needed to use twice the resolution as in the previous section in order to avoid spurious peaks in $\mathcal{M}_{\rm b}$ at small radii.

\subsection{Elevated Bases}

\begin{figure*}
    \centering
    \includegraphics[width=\linewidth]{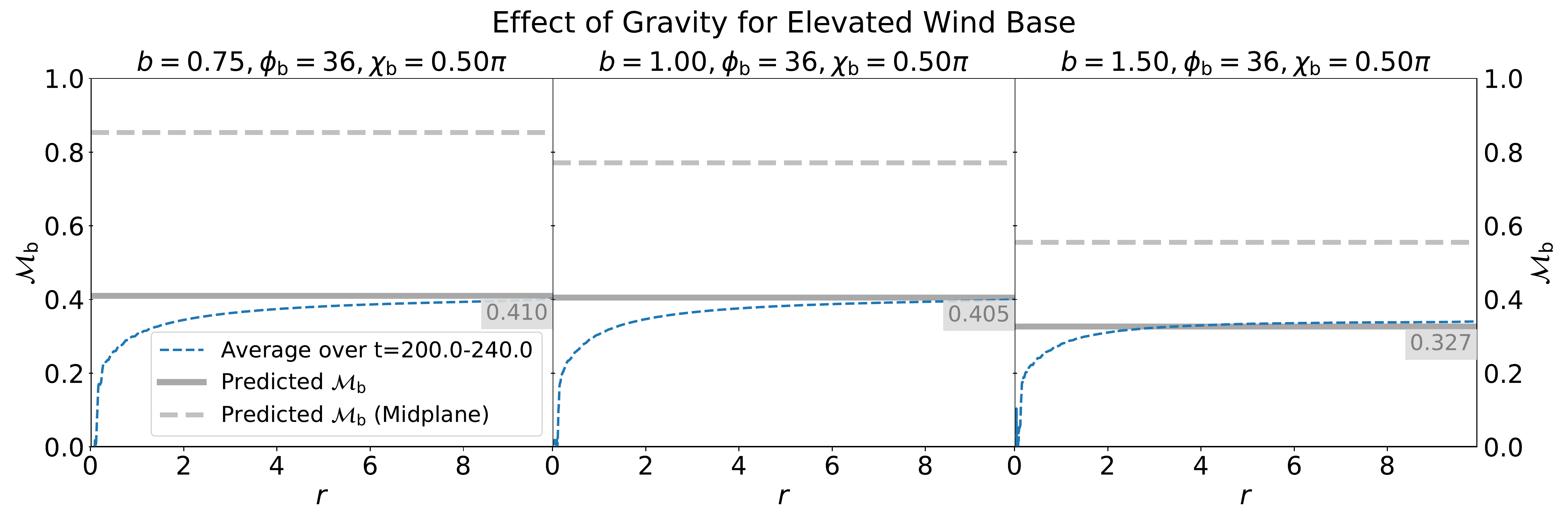}
    \caption{
    Comparison of the launch Mach numbers $\mathcal{M}_{\rm b}$ as a function of radius (in units of the gravitational radius) for winds (with gravity/centrifugal forces included) launched perpendicularly from a base elevated by $\phi_{\rm b}=36^\circ$ with density power law slopes $b=0.75,1.00,1.50$ from left to right. The blue dashed lines indicate the output from hydrodynamic simulations averaged over a range of times to smooth the effect of oscillations since the solutions are not perfectly steady, especially at small radii. The grey bands are as in Figure \ref{fig:ss_geometry_Mb}. Note the switch to a linear x-axis scale as per \citealt{Clarke_2016} since the solution is no longer scale-free, allowing us to highlight the region over which the self-similar solution is a good approximation.
    }
    \label{fig:nss_p36_Mb}
\end{figure*}

\begin{figure*}
    \centering
    \includegraphics[width=\linewidth]{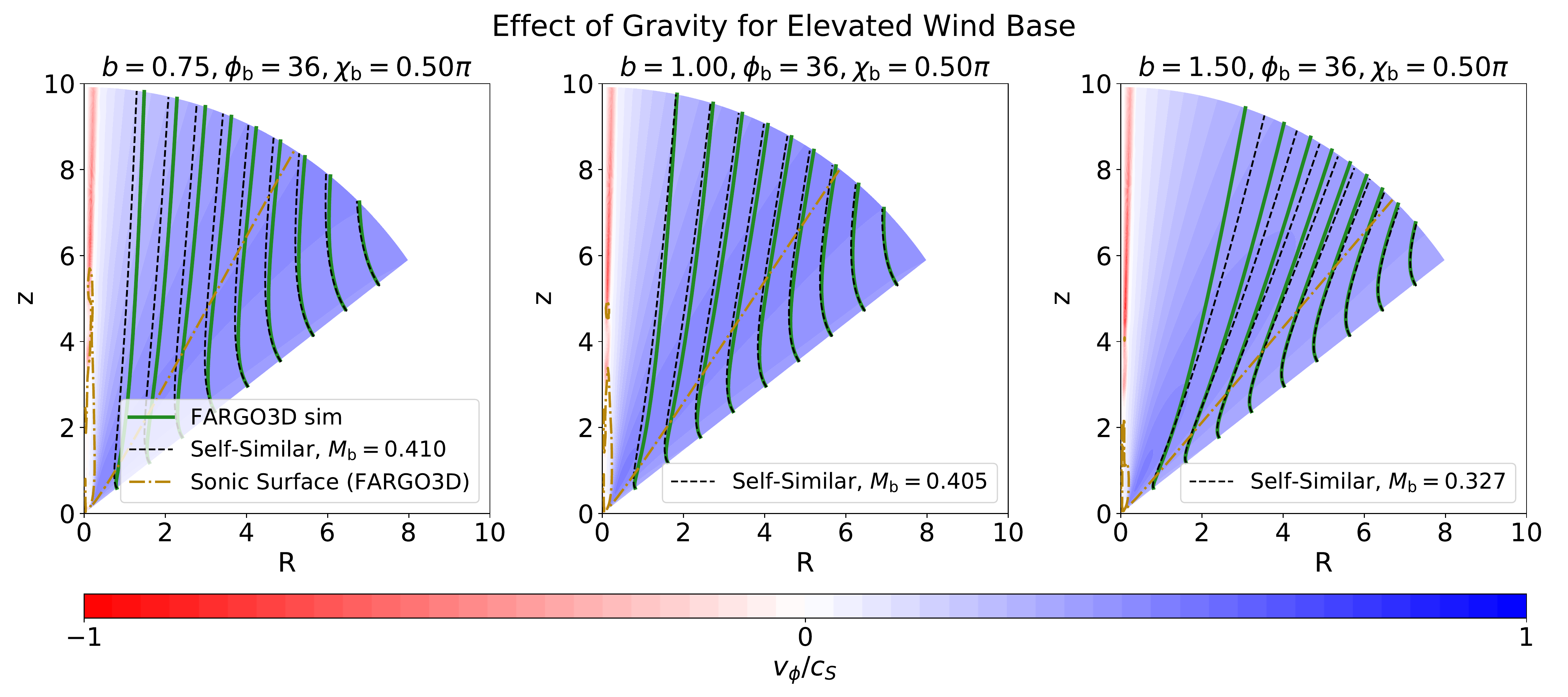}
    \caption{As with Figure \ref{fig:ss_geometry_streamlines} but for the wind models in Figure \ref{fig:nss_p36_Mb} with gravity/centrifugal forces included. The hydrodynamic simulation outputs have been averaged over the same range of times as Figure \ref{fig:nss_p36_Mb}. The self-similar solutions are still very decent representations of the streamline morphology.}
    \label{fig:nss_p36_streamlines}
\end{figure*}

Figure \ref{fig:nss_p36_Mb} and \ref{fig:nss_p36_streamlines} show the launch Mach numbers and streamlines respectively for elevated wind bases with $\phi_{\rm b}=36^\circ$. In all cases $\chi_b=\pi/2$. The outputs of the simulations are averaged over a range of times in order to average out small fluctuations.

Figure \ref{fig:nss_p36_Mb} illustrates that as found by \citet{Clarke_2016} the launch Mach numbers are roughly constant in the outer disc where $r>r_G$ (i.e. $r>1$ on the plots) with values that are well-predicted by $\mathcal{M}_{\rm b, max}$.
\citet{Clarke_2016} argued that the curvature dominates over gravity/centrifugal forces when $(r/r_G) \times r/r_{\rm eff} \gg 1$; consequently, we see the closest agreement in the $b=1.5$ cases where the wind is most strongly curved at the base.
Moreover material continues to be launched somewhat inside the gravitational radius, albeit more slowly \citep{Font_2004,Clarke_2016}.

The self-similar streamlines continue to provide a good model for the morphology of the streamlines in Figure \ref{fig:nss_p36_streamlines}, especially in regions of high curvature.
Equation \ref{eq:u_azimuthal} strictly balances centrifugal force with gravity at the midplane. As $z$ is increased at the elevated wind base, gravity weakens but the centrifugal force is not affected. Thus centrifugal force dominates over gravity at our wind base, resulting in a net outward force.
For the low $b$ winds, the streamline curvature is small so this net force has a more significant effect and pushes the streamlines to a larger $R(z)$.
In the $b=1.5$ case this effect is subdominant to the existing curvature. Instead in this case, since there is no flow from small $r_{\rm b}$, the region near the z-axis is poorly supplied with material. Thus, there are much stronger density gradients in the $\phi$ direction, which cause the streamlines to curl upwards more strongly to fill the spatial domain.

In the simulations described above, we use boundary conditions designed to impose $u_r=0$ such that the winds should launch perpendicularly.
However just above the base, the flow develops non-zero $u_r$ due to the streamline curvature. 
As $\mathcal{M}_{\rm b}\to0$ for $r \ll r_G$, then at small radii, the angle the wind makes with the base $\chi_{\rm b} \to 0$ also.
By $r\approx0.2~r_G$, the launch velocity drops sufficiently that (when measured just above the wind base) $\chi_{\rm b}\approx\pi/4$ which we deem sufficient to explain the non-perpendicular streamlines in the inner disc in the simulations of e.g. \citet{Picogna_2019}. For this reason, we do not further impose a non-perpendicular launch.

\subsection{Radial Temperature Profiles}
\begin{figure*}
    \centering
    \includegraphics[width=\linewidth]{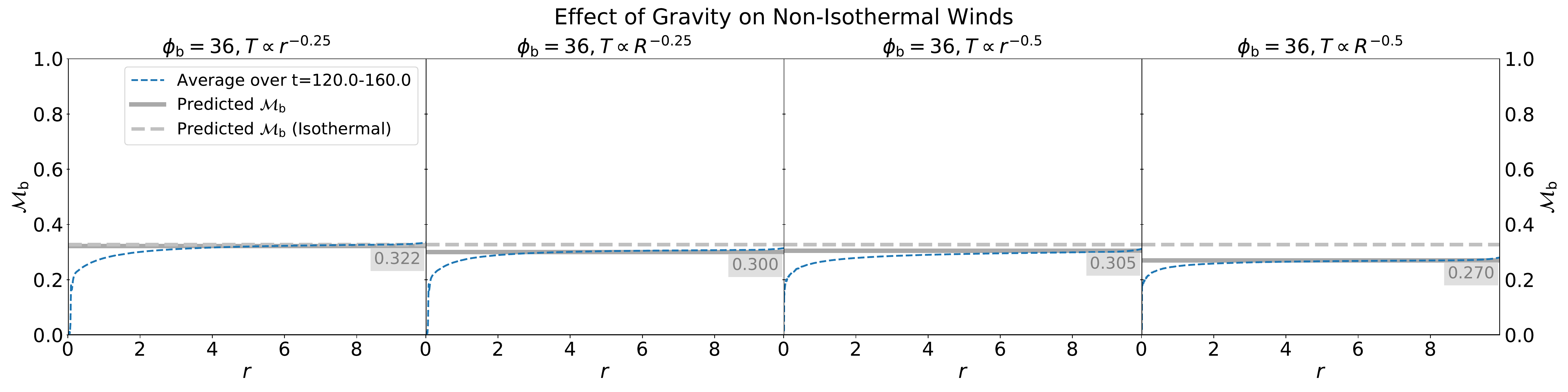}
    \caption{As with Figure \ref{fig:niT_Mb} but including gravity/centrifugal forces (hence $r$ is in units of the gravitational radius). The blue dashed lines indicate the output from hydrodynamic simulations averaged over a range of times. The grey bands are as in Figure \ref{fig:niT_Mb}.}
    \label{fig:full_t025_Mb}
\end{figure*}

We now present our most complete models by reintroducing non-isothermal effects. Since material is not launched from the inner grid radius anyway, it should not matter which boundaries conditions are applied, so for consistency we use a constant velocity across the boundaries, rather than constant Mach number. We plot the Mach number at the base as a function of radius in Figures \ref{fig:full_t025_Mb}.

The resulting $\mathcal{M}_{\rm b}$ profiles are consistent with the results found in previous sections.
The effect of gravity is still to stifle mass loss at small radii.
However, when compared to the isothermal case, the wind is a little more readily launched from smaller radii the larger the temperature gradient. This is because although the gravitational radius has been fixed, the ratio of thermal energy to gravitational energy declines more slowly (as $r_b^{1-\tau}$) with decreasing radius due to the higher temperatures at smaller radii so there is more thermal energy to drive the wind.
Moreover, the reduced launched velocities mean that the radius of curvature is smaller, further pushing us into the regime where $(r/r_G) \times r/r_{\rm eff} \gg 1$ \citep{Clarke_2016}. Consequently, the profiles of $\mathcal{M}_{\rm b}$ are very flat, and in good agreement with the predicted values for non-isothermal winds (Table \ref{tab:Mb_niT}).

\begin{figure*}
    \centering
    \includegraphics[width=\linewidth]{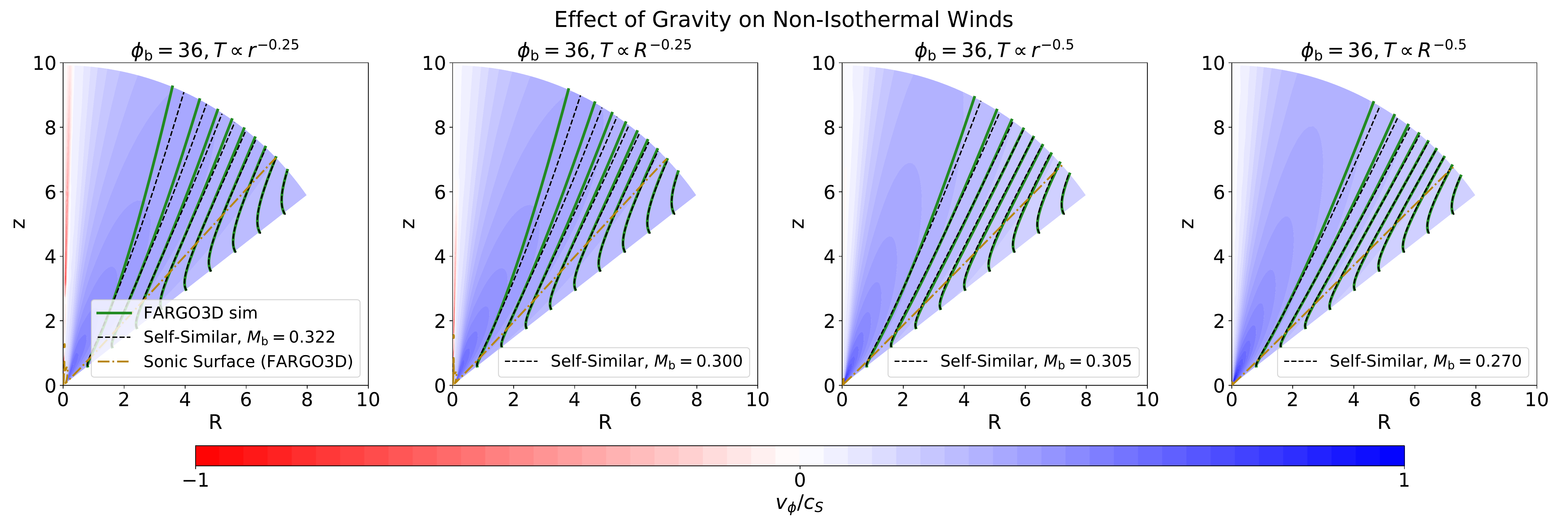}
    \caption{As in the $b=1.5$ panel of Figure \ref{fig:ss_geometry_streamlines}, a wind launched from $\phi_{\rm b}=36^\circ$ but including gravity, centrifugal force and temperature gradients of varying steepness and direction. Once again, the self-similar solutions are still very decent representations of the streamline morphology.}
    \label{fig:full_streamlines}
\end{figure*}

Thus, while the temperature gradients do lower the launch velocities, they also have the effect of mitigating against gravity and centrifugal force, thus reducing the deviations from self-similarity.
This can also be seen in the streamlines, which we plot in Figure \ref{fig:full_streamlines}. In particular, comparing the two left-most panels ($\tau=0.25$) and the two right-most panels ($\tau=0.5$) the steeper temperature gradients, have closer agreement between the streamlines in the hydrodynamic simulations and the self-similar models. In this case, even the deviations at large radii are less apparent, because the additional thermal energy assisting the launch at small radii means that the region near the z-axis is no longer slightly inflowing and hence not so depleted of material. Again, whether the temperature depends on the spherical radius (first and third panels) or cylindrical radius (second and fourth panels) has no bearings on the results.

Therefore, we conclude that the effects of gravity/centrifugal forces do not strongly modify the launch velocities or streamlines at suitably large radii compared to $r_G$ for elevated wind bases.
This behaviour is only reinforced by the presence of radial temperature gradients.
At small radii, $\lesssim r_G$, it can become harder to drive an outflow and the launch velocities are lowered; correspondingly the winds no longer launch quite perpendicularly (note that this effect is oppositely directed to the scale-free case where non-perpendicular launches act to raise the Mach number at the base).
This does not impact on the validity of the self-similar solution near the base at radii beyond the gravitational radius, but weakened flow from small radii can result in a stronger upward curvature at large radii in the case of weak temperature gradients.
Thus as found by \citet{Clarke_2016}, the self-similar solutions have general applicability to describe the launch velocity and streamline structure of thermal winds when gravitational and centrifugal forces are included and this agreement is only strengthened by introducing a moderate temperature gradient (see third and fourth panels of Figure \ref{fig:full_streamlines}).

\section{Double Power Laws}
\label{sec:dbl_pow}
Our final consideration is that a power law of infinite extent will never completely describe the density in the upper layers of discs.
Most simply, at some point, there must be a cut-off at the edge of the disc.
Moreover, the density structure at the wind base is dependent on the mechanics of the heating; for example \citet{Hollenbach_1994} argued that the power law slope transitions from $1.5$ at $r<r_G$ to $2.5$ for $r>r_G$ as the origin of the dominant diffuse ionising photon flux transitions from material directly above the wind base to material at $r_G$\footnote{However in hydrodyamical simulations that do not impose a base profile, such a transition to a steep $b>2$ power lw does not apparently occur at $r=r_G$ \citep{Wang_2017,Yorke_1995}}.
Regardless of the reason for the transition, if a disc is described by different power laws at small/large radii, we expect the appearance of the transition radius $r_{\rm t}$ to break self-similarity, and this may affect the applicability of the self-similar solutions.

To see how these deviations from self-similarity manifest, we consider double power laws both of the form used by \citet{Font_2004}, and an inverse equivalent:

\begin{equation}
    \rho \propto \left(\frac{2}{{r/r_{\rm t}}^{5b_1}+{r/r_{\rm t}}^{5b_2}} \right)^{1/5}
    \label{eq:dblpow},
\end{equation}
\begin{equation}
    \rho \propto \left(\frac{{r/r_{\rm t}}^{-5b_1}+{r/r_{\rm t}}^{-5b_2}}{2} \right)^{1/5}
    \label{eq:invpow}.
\end{equation}
For clarity we will always choose $b_2>b_1$; then the former of these profiles is used for a profile that transitions to a steeper power law at large radii, whereas the latter transitions to a shallower power law.

We consider the following three combinations of $b_{\rm in}$ and $b_{\rm out}$: $1.50/1.75$, $1.75/1.50$, $1.50/2.50$.
In the first two cases (power law transitions to $b<2$), 
we expect that on their own, both the inner and outer regions of the disc could launch a self-similar wind. In the latter case (power law transitions to $b>2$),
we would not expect a self-similar solution to exist for the outer disc.
That said, a single power law with $b=2.5$ may still permit a \textit{non-self-similar} wind solution \citep{Font_2004}; in particular the requirement for such a case to be diverging faster than spherical may be circumvented by suppressing the contribution from streamlines with small $r_{\rm b}$; contributions to this in \cite{Font_2004} include gravity impeding the launch for $r_{\rm b}<r_G$, and their use of a reflecting inner boundary condition which prevents material launched from $r_{\rm b}<r_{\rm in}$ entering the simulation domain.

In each case, we will assume $r_{\rm t}=1$: firstly without gravity (in which case there is no physical significance to this value), and with gravity for the case with $b_2=2.5$ (in which case we use units of $r_G$ such that $r_{\rm t}=r_G$).
We thus use a grid that now extends from $r=0.01$ to $r=100$, in order to have an equal two decades in radius either side of the transition. Moreover, two decades is enough to ensure that the slope of the density profile at the edges of the grid should differ from the relevant limiting value by less than 1 per cent of difference in the $b$ values in all cases.

\subsection{Power Law Transitions to $b<2$}
\begin{figure}
    \centering
    \includegraphics[width=\linewidth]{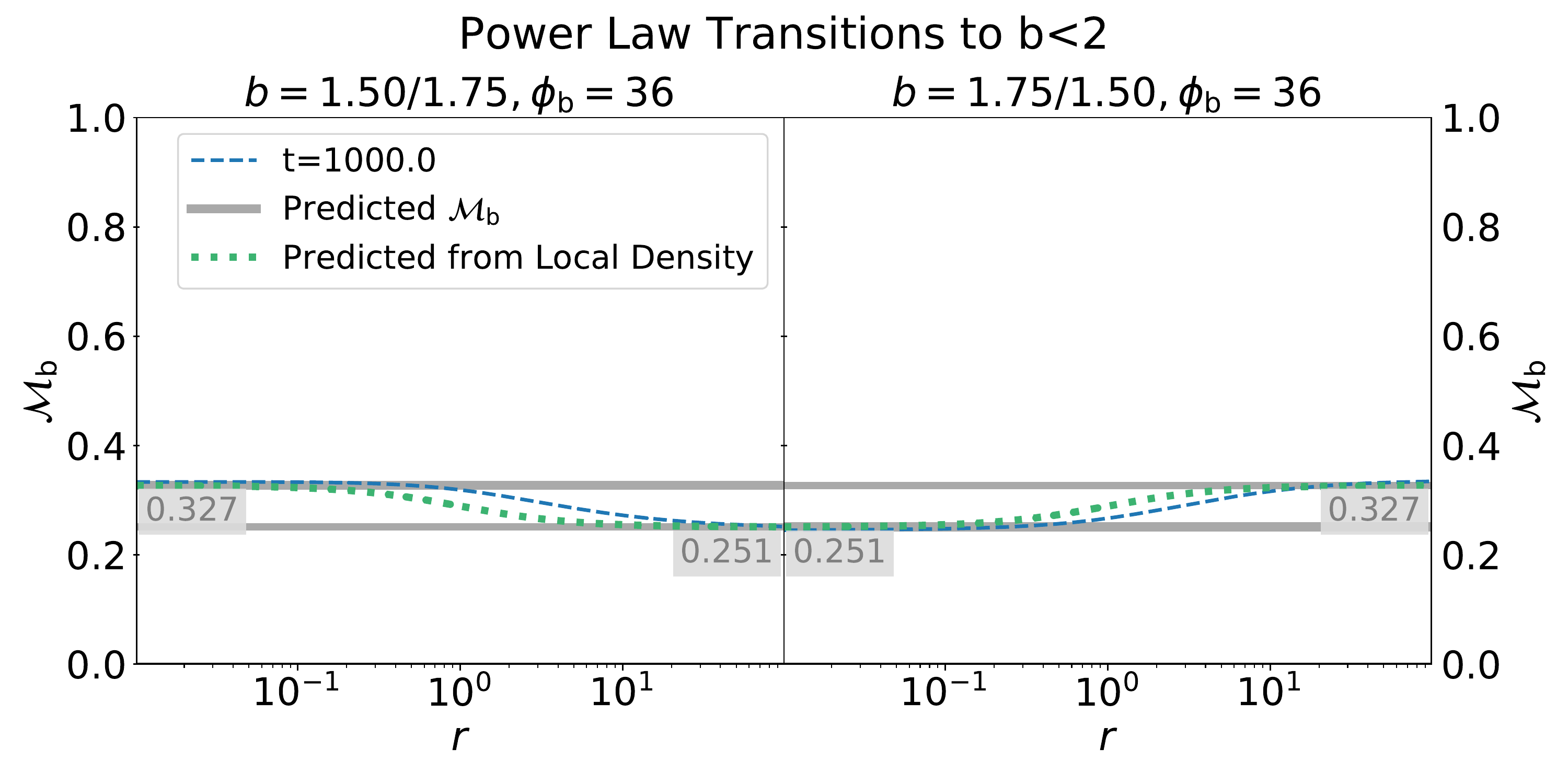}
    \caption{As with Figure \ref{fig:ss_geometry_Mb} but for a double power law density profile with $b_1=1.5$, $b_2=1.75$ in both the steepening and flattening cases. The two dark grey bands now represent the expected value of the Mach number for a self-similar wind with $b=1.5$ or $b=1.75$. In addition, the green dotted line shows the Mach number that would be expected for the local value of the density gradient $b_{\rm eff} = \frac{\partial \ln(\rho_{\rm b})}{\partial \ln(r_{\rm b})}$.}
    \label{fig:b150_b175_Mb}
\end{figure}

First, we consider cases where in either extreme, the density profile would permit a self-similar wind model, i.e. $b_1,b_2<2$. We consider two cases, one that scales like $r^{-1.5}$ in the inner disc and $r^{-1.75}$ in the outer disc (steepening case), and one where these values are reversed (flattening case). The Mach numbers at the base are shown in Figure \ref{fig:b150_b175_Mb} for the elevated base with $\phi_{\rm b}=36^\circ$.

In both the steepening and flattening cases, Figure \ref{fig:b150_b175_Mb} shows that for radii $r\lesssim1$ the launch Mach numbers (blue lines) are those that we would expect given the density gradient at these radii. Beyond $r=1$, the velocities smoothly transition towards the value appropriate for the outer disc and appear to be asymptoting to that value at $r\gtrsim30$.
Therefore, regardless of whether the inner or the outer disc is the steeper, there is an asymmetric behaviour about the transition point between the two regimes of the density profile, with the launch velocity relatively unaffected within the inner disc, but the outer disc feeling the effects over 1 to 2 orders of magnitude. 

In previous sections we have shown that the launch velocity is mainly set by the value of b and the elevation of the wind base.
Thus, we also show the Mach number that would be expected for the local value of the density gradient $b_{\rm eff} = \frac{\partial \ln(\rho_{\rm b})}{\partial \ln(r_{\rm b})}$ as the green dotted line. This shows a very similar shape to the Mach numbers that result in the hydrodynamic simulation, but simulation profile is shifted to higher radii by a factor of roughly $3-4$.
Since this wind flows from smaller to larger radii, this is consistent with a picture of outwardly-directed causality, where 
the outer disc retains some memory of the launch velocity appropriate to the inner disc over a significant radial range. 

Thus a steep density gradient at small radii, which launches a wind more slowly, can suppress launch velocities at intermediate radii.
We note the similarities between this scenario, where the inner wind is more curved and is effectively acting to constrain the streamlines in the outer disc somewhat, with the restriction of the domain we performed in Section \ref{sec:domainfilling}.
However, in each extreme, particularly in the inner disc, the self-similar model is still applicable.

\subsection{Power Law Transitions to $b>2$}
\label{sec:dblpow_2}
Now we consider a base density profile described by equation \ref{eq:dblpow} with $b_1=1.5$ and $b_2=2.5$, i.e. the same case studied by \citet{Font_2004} following \citet{Hollenbach_1994}.
This scenario has a density gradient in the inner disc which would be suitable for being described by a self-similar model.
In the outer disc, the density gradient is too steep for our self-similar models, but as discussed above, may still permit a wind to be launched; \citet{Font_2004} find there is some contribution to the wind from this region, though most of the mass is lost from inside the transition.
We consider both the case without gravity and a case where the transition radius $r_{\rm t}$ corresponds to $r_G$. Note that the former case may be seen as equivalent to the extreme limit of $r_{\rm t} \gg r_G$. 

\begin{figure*}
    \centering
    \includegraphics[width=\linewidth]{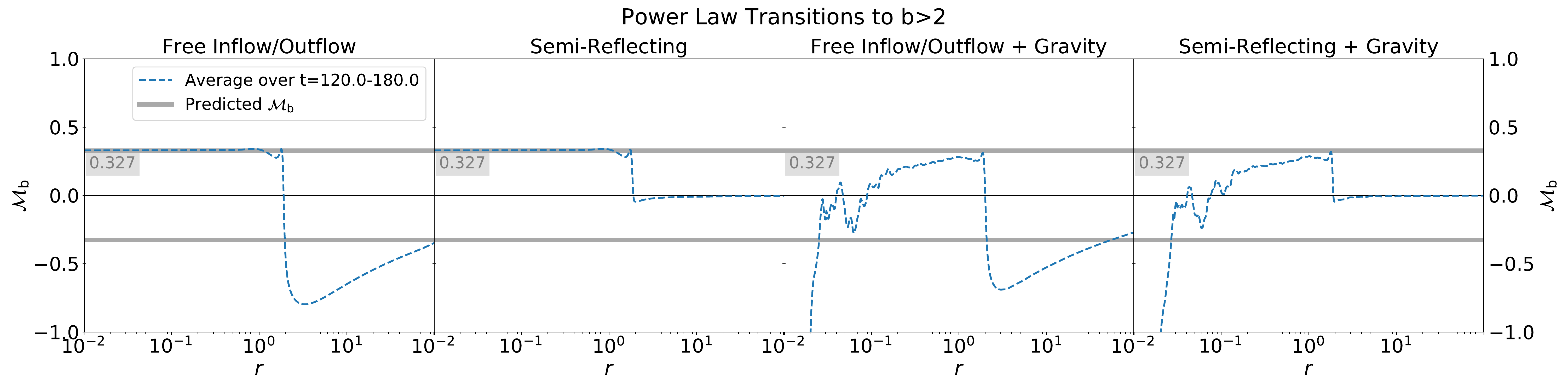}
    \caption{Comparison of the launch Mach numbers $\mathcal{M}_{\rm b}$ for a double power law density profile with $b_1=1.5$, $b_2=2.5$ launched from $\phi_{\rm b}=36^\circ$. The odd panels have a free outflow/inflow at the launch plane while the even panels have a reflecting boundary to prevent inflow for $r>1.9 r_G$.
    The darker grey band represents the predicted $\mathcal{M}_{\rm b, max}$ from the self-similar models with $b=1.5$ and the grey label shows its value.
    }
    \label{fig:b150_b250_Mb}
\end{figure*}

We investigate this scenario using two sets of boundary conditions.
First we allow for a free perpendicular inflow or outflow at the launch plane.
The Mach numbers at the base are shown in the left most panel of Figure \ref{fig:b150_b250_Mb} for the case with no gravity. We see that at small radii $r<r_G$, the flow adopts the expected value from the self-similar solution. This is in line with the two-regime winds studied above where the inner disc is not strongly affected by the outer disc.
The transition from $b=1.5$ to $b=2.5$ is centred on $r=1$ (at which location $b_{\rm eff} \equiv -\frac{\partial\ln\rho_{\rm b}}{\partial\ln r_{\rm b}} = 2$); in the vicinity of this point, the flow feels the effect of the more strongly declining density and begins to deviate from self-similarity before dropping off rapidly.
Thus, at radii $r \gtrsim 1.9$, the velocities change sign and for there is a flow back into the launch plane. While this flow is subsonic, its velocities are not insignificant and are on the order of the sound speed.

Since we do not model the underlying disc, it is unclear how much resistance this might provide to such a back flow.
We therefore also consider the extreme opposite situation (such that our two boundary conditions ought to bracket the "true" behaviour), where 
for $r > 1.9$\footnote{We tested other locations in the range $1 \leq r \leq 2$ for the change in boundary conditions and found that $r= 1.9$ reduced any overshooting of the velocities near the change.}, the launch plane becomes reflecting to prevent this flow back into the disc.
The resulting Mach numbers for this "semi-reflecting" setup are shown in the second panel of Figure \ref{fig:b150_b250_Mb}; the inner disc is unaffected by this change in boundary conditions while in the outer disc the flow instead becomes radial along the disc surface.
We note that the small velocities in the outer region are not perfectly steady but the oscillations are very small and average out smoothly. We also attempted to set the boundaries self-consistently using a diode based on the sign of the velocities but found this was unstable and showed large deviations from the steady state.

Having established that a wind is only possible within $r \lesssim r_{\rm t}$, we reran the models with both boundary conditions including gravity and centrifugal force - the Mach numbers are shown in the third and fourth panels of Figure \ref{fig:b150_b250_Mb} respectively. We again set $r_{\rm t}=r_G=1$ in line with \citet{Hollenbach_1994}.
In the inner disc, the Mach numbers greatly resemble the trends observed in the case of a single $b=1.5$ power law when gravity is included (right hand panel, Figure \ref{fig:nss_p36_Mb}, in that there is no outflow inside of $r\approx0.1$, and the velocities rise in the range $0.1-1.0$, beginning to flatten off as $r\to1$ though not quite yet reaching the self-similar value of $0.327$.  Beyond $r\approx 2$, the behaviour appears identical to the left-hand panels of Figure \ref{fig:b150_b250_Mb} and we find either a strong and smooth flow back into the disc, or no flow depending on our boundaries. Thus, we find that where $b>2$, winds only originate from a very limited range of radii in the vicinity of $r_{\rm t}$, and centrifugal forces do not assist the launch of such a wind.
This is in agreement with the results of \citet{Font_2004} who found that most of the mass-loss in such a scenario comes from $0.1 < r/r_G < 2$.

\begin{figure*}
    \centering
    \includegraphics[width=\linewidth]{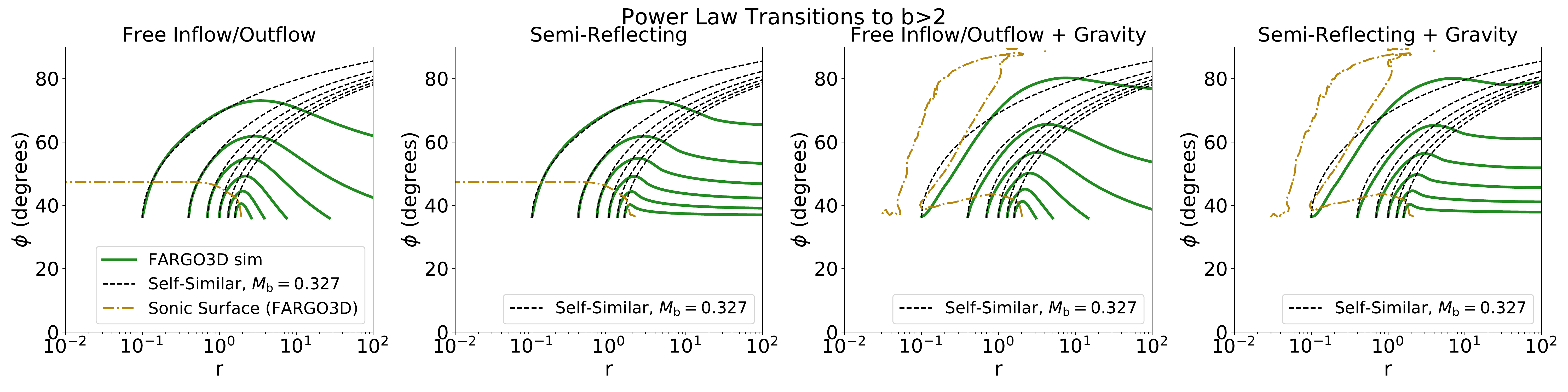}
    \caption{The elevation $\phi$ (degrees) of the streamlines as a function of radius for the `single-regime' models shown in Figure \ref{fig:b150_b250_Mb}. The streamlines are shown for $r_{\rm b}=0.1-1.6$ in steps of $0.3$. It is apparent that due to the steep density drop-off, larger deviations between the self-similar solutions and the \textsc{FARGO3D} results are present than for winds with base densities described by single power laws as the wind curves back towards its base. The gold dot-dashed line shows the sonic surface, demonstrating that the wind remains supersonic in this region at $r \gtrsim r_G$; hence the material here is unbound.}
    \label{fig:b150_b250_streamlines}
\end{figure*}

The large-scale morphology of these winds would be expected to be very different from the self-similar models since either a radial flow or a flow back into the disc occurs at large radii. This means a lack of supply of material to support the streamlines which instead curl back towards the base. To emphasise this behaviour, Figure \ref{fig:b150_b250_streamlines} shows the streamlines in polar coordinates i.e. $\phi$ as a function of $r$.

For the models without gravity, the innermost parts of the streamlines near their bases agree well with the self-similar models and the morphology is independent of the boundary condition choice at larger radii. Somewhere near $r=1$, they begin to flatten off and the elevation peaks in the range $1 \lesssim r \lesssim 3$. Beyond this point there are two possible fates. In the case of free outflow, the elevation decays - most rapidly for the lower streamlines and more slowly for the higher streamlines, until they reintercept the launch plane. When the boundaries are reflecting, the streamlines cannot cross it, and the material following them remains in the grid. This provides an upwards pressure gradient, self-consistently reducing the downward velocities of the wind, resulting in the elevation of each streamline levelling off again such that they become radial as expected. A whole range of asymptotic elevations are possible, so there is no particular opening angle of the wind and not all material returns near to the base. This means that there are no particularly evacuated regions of density, moreover at any radius, we found the density contours to be roughly spherical in agreement with \citet{Font_2004}. In both cases the streamlines are concave upwards. Despite these differences close to the midplane, the qualitative picture of declining $\phi$ holds for both sets of the boundaries, thus we expect to see streamlines curling back towards the disc regardless of where the correct boundary conditions lie relative to the two extremes shown here and hence we are confident that the "true" behaviour is reasonably well captured by these models.

The polar streamline plots in the presence of gravity (right hand two panels of Figure \ref{fig:b150_b250_streamlines} illustrate that as seen in the launch velocities, the large scale morphology at $r>r_G$ is little affected by gravity.
The streamlines do reach slightly higher elevations - analogously to the upward curling seen at large radii in the $b=1.5$ models with gravity.
Within the transition radius the streamlines follow the self-similar solution for launch radii $\gtrsim 0.3 r_G$.

In the particular example given, where $r_{\rm t}=r_G$, this implies that the self-similar solution is only approximately valid over a factor $3-10$ in radius around $r_{\rm t}=r_G$, with deviations at small radii caused by the role of gravity and at large radii by the transition to a steeper density profile. More generally, the self-similar solutions can describe the region $r_G \lesssim r \lesssim r_{\rm t}$; we emphasise that although \citep{Hollenbach_1994} argued that $r_{\rm t}=r_G$ for EUV-driven photoevaporative winds, the location of any turnover in the density profile may be sensitive to details of the radiation transfer such as the roles of direct/diffuse radiation field (dependent on shielding/scattering processes). Radiation hydrodynamic simulations \citep[e.g.][]{Richling_1997, Wang_2017} suggest that $r_{\rm t} > r_G$ and hence a larger role for self-similar solutions in describing EUV-driven winds than our results would suggest.

\section{Discussion}
\label{sec:discussion}
The derivation of the self-similar solutions and their comparison to hydrodynamic simulations is motivated by both theoretical and observational considerations.
Theoretically, they form a robust toy model for more complicated scenarios, which allows us to better understand what determines the launch velocities of photoevaporative winds - which contribute to setting the mass loss profiles found by radiation hydrodynamic models - and how these are affected by different forces acting on the wind.
Observationally, their relative ease of application allows us to interpret, for example, forbidden line spectra without needing dedicated radiation hydrodynamic simulations of photoevaporation \citep{Ballabio_2020}.

\subsection{Comparison to Photoevaporation Simulations}
We have demonstrated that the effects of gravity and centrifugal force make only a small impact on the launch velocity and streamline morphology of thermal winds
and that radial temperature gradients cannot decrease $\mathcal{M}_{\rm b}$ by much more than $\sim10$ per cent compared to the isothermal case.
The key determinant of the launch Mach number is the elevation of the wind base, combined with the requirement of the asymptotic morphology of the streamlines at large radii to be space-filling, and our simple hydrodynamic simulations follow these predictions.
Therefore, as long as, for example, the wind base isn't too strongly flared, we should reasonably expect more detailed radiation hydrodynamic simulations such as those of \citet{Wang_2017,Picogna_2019} to show velocities that agree with the self-similar models.

\cite{Wang_2017} find that the wind region is situated above $z/R\approx0.6$, corresponding to a launch plane at $\phi_{\rm b} \approx 30^\circ$. By inspecting Figure \ref{fig:Mb_contours} we see that this corresponds to a Mach number $\mathcal{M}_{\rm b} \lesssim 0.4$ assuming $b\approx1.5$ (as for the \citet{Hollenbach_1994} wind model at $r \leq r_G$
).
In their Figure 2, \citet{Wang_2017} also show the Mach number as a function of distance along streamlines originating from $R=5$ and $R=15$ au, which correspond to around 0.5 and 1.5 gravitational radii (equation \ref{eq:r_G}). In both cases, at the base the Mach number appears to be tending towards a value in the range $0.3-0.4$ in good agreement with our estimate. This is despite their simulations suggesting a significant role for adiabatic cooling in the thermodynamics of the wind.

On the other hand, the simulations of \citep{Picogna_2019} appear to agree with these predictions less well. Their wind base is well-fitted by $\phi_{\rm b}=36^\circ$, with $\rho \propto r^{-1.5}$ but the average Mach number here is more like $\langle\mathcal{M}_{\rm b}\rangle=0.1$ across a wide range of radii (R. Franz, Private Communication), whereas we would expect a value of $0.327$. However, the Mach numbers quickly rise to around $0.3$ within a couple of degrees of the base, much faster than do the self-similar solutions. This is likely due to the impact of their definition of the base lying in a region of steep vertical temperature gradient normal to the base; above the base as the temperature gradients become smaller, the solution reverts to what we would expect would be appropriate to fill the domain.

While thermal winds can also be driven from protoplanetary discs by an external source of FUV radiation, these have a larger gravitational radius such that in the usual, "subcritical" regime, the mass loss is dominated by the least bound material at the disc edge \citep[e.g.][]{Adams_2004,Haworth_2019}. This strong dependence on a characteristic radius means our self-similar solutions have limited applicability to \textit{externally} photoevaporating discs.

\subsection{Disc Clearing}
Inside-out clearing of protoplanetary discs is backed by observations such as discs with inner cavities - many with negligible accretion rates \citep{Owen_2012b} - and the distribution of discs in colour-colour diagrams \citep{Koepferl_2013}.
It naturally occurs when the mass loss rates $\dot{\Sigma} \propto \rho_{\rm b} u_{\rm b}$ ($\propto r^{-(b+\tau)}$ in the scale-free models), decline faster than the disc surface density $\Sigma$ (often assumed to be $\propto r^{-p}$; $p=1$ and $p=1.5$ are popular models). This minimises the depletion timescale near the critical radius, inside of which gravity impedes launching a wind. A gap opens and allows the inner disc to drain rapidly, while photoevaporation continues to erode the outer disc from the inside-out \citep{Clarke_2001,Alexander_2006b}.

Since our `two-regime' models suggest that sufficiently outside a density transition, the wind can forget the inner disc conditions and become approximately self-similar, we might expect even in this case, the self-similar solution to be a good approximation. We caution that a severely depleted inner disc is a much stronger deviation from the density profile than considered here. However, \citet{Alexander_2006a} do note that - while fluctuating considerably - there was no evidence for a strong radial dependence in the launch velocities of their directly EUV-irradiated discs, a feature we see repeatedly in our models as a hallmark of self-similarity. 
This result is thus useful if we wish to apply self-similar solutions to clearing discs. This may be when thermal winds are most important: \citet{Pascucci_2020} suggest an evolutionary picture where in full discs, an inner MHD wind blocks the radiation required to launch a thermal wind from the outer disc thus preventing such winds for much of the disc's lifetime, while an outer thermal wind may be present for discs with an inner hole that are likely in the process of clearing.

\subsection{Implications for Line Profiles from Discs}
It is also worth considering what, if any, effect our more generalised self-similar models would have on the inference of wind properties from blueshifted forbidden emission line profiles, such as in the work by \citet{Ballabio_2020}. In that work they found that the peak and centroid velocities of the blueshifted lines were most strongly determined by the sound speed, with the density gradient and normalisation less important. On the other hand, when it came to the Full Width at Half Maximum (FWHM) of the lines, not only was the sound speed important, but if there was more emission from small radii - as in cases with higher $b$ or lower $\dot{M}_{\rm w}$ (such that the point where the density exceeds the critical density for producing forbidden lines moved inwards) - then there would be additional emission at large blueshifts, increasing the line width.

\begin{figure*}
    \centering
    \includegraphics[width=0.80\linewidth]{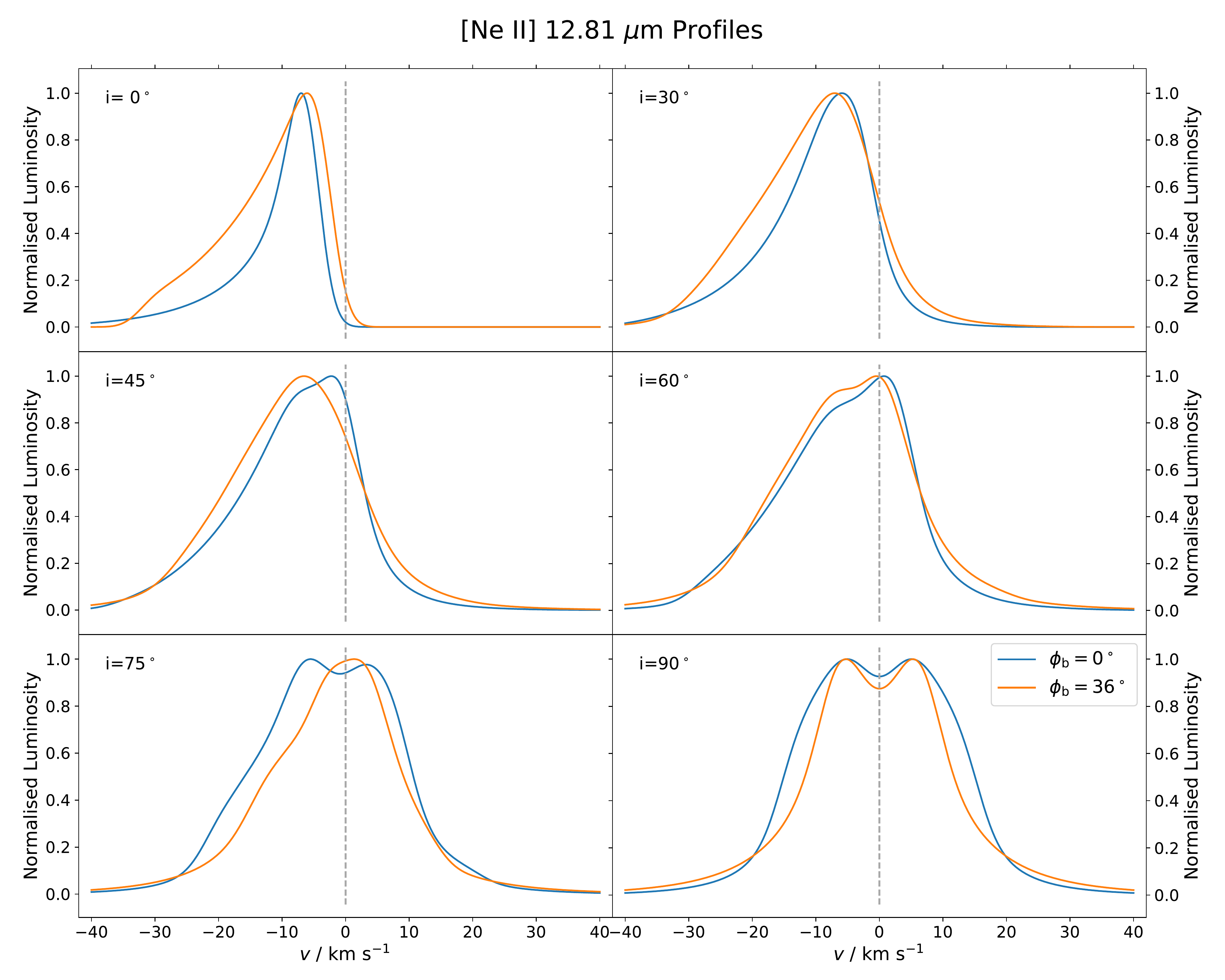}    \caption{Comparison of the [\ion{Ne}{II}] $12.81~\mu m$ line profiles for self-similar winds launched from $\phi_{\rm b}=0^\circ$ (blue) and $\phi_{\rm b} = 36^\circ$ (orange)} at various inclinations. The profiles have each been normalised by their maximum value.
    \label{fig:line_profiles}
\end{figure*}

While \citet{Pascucci_2011,Ballabio_2020} noted that launch from an elevated base can reduce the blueshift of the line peak by $1~\mathrm{km~s^{-1}}$, we present in Figure \ref{fig:line_profiles} emission profiles for the [\ion{Ne}{II}] $12.81~\mu\mathrm{m}$ line at different viewing inclinations in order to compare directly winds launched from the midplane with those launched from $\phi_{\rm b}=36^\circ$.
We followed the equations used in \citet{Ballabio_2020} and used the same set of atomic constant for the line transition \citep{Glassgold_1997}. Briefly, the level populations for NeII are calculated with a density dependent prescription taking into account the critical density for the transition \citep{Glassgold_1997} and the optically thin line profile includes both thermal broadening and Doppler shift due to the local line of sight velocity.
We calculate the emission on a spherical grid spanning radii $r=[0.03,10] r_G$ and elevations $\phi=\pm[\phi_{\rm b}, 75^\circ]$. The density at $(r,\phi) = (r_G, \phi_{\rm b})$ is normalised to the same number density $n_G = 2.8 \times 10^4 ~\mathrm{cm}^{-3}$.
We assume that in the disc, dust provides the dominant source of opacity and blocks the receding portion of the wind, which is consistent with observations of net blueshifted [\ion{Ne}{II}] emission arising exterior to dust cavities \citep{Pascucci_2011,Pascucci_2020}. Therefore for simplicity, we assume that the disc midplane is infinitely opaque, and all other material is optically thin, when determining which parts of the wind are visible at any inclination.

The most striking effect is on the FWHM: the width of the line when the base is elevated is greater for discs which are viewed face on, but the lines become narrower at high inclinations; this largely wipes out the dependence of the FWHM on viewing inclination predicted for winds launched from the disc mid-plane \citep{Ballabio_2020}. This occurs since the morphology of the streamlines is more vertical when the base is elevated (e.g. Figure \ref{fig:streamline_comparison_analytic}), meaning that near the base the velocity vectors are directed more along the polar direction ($i=0^\circ$) than towards high inclinations ($i\lesssim90^\circ$).
Moreover, we note that our profiles are normalised; the absolute luminosities in the line wings are relatively insensitive to the launch height, but there is a strong loss of luminosity near zero velocity when the wind launches from an elevated base.
This results from a lack of emission close to the disc mid-plane which no longer hosts an ionised wind when the wind base is elevated.
Whereas, near the polar axis and at large radii in the wind, where the highest velocity material exists, the absolute velocity is insensitive to the launch speed at the base - and hence its elevation - so the line wings are relatively unaffected, especially for face on discs when this material flows along the line of sight.
Although the resulting differences are only a few $\mathrm{km~s^{-1}}$, the error bars on the [\ion{Ne}{II}] FWHM data used by \citet{Ballabio_2020} tend to be comfortably $\lesssim 5~\mathrm{km~s^{-1}}$, so this should be taken into account in future for a more accurate interpretation of the FWHM as a function of inclination, particularly if the aim is to discriminate between sound speeds which differ between EUV X-ray winds by a few $\mathrm{km~s^{-1}}$.

Since the self-similar models are applicable near the base even in the presence of gravity/centrifugal forces, or double power law density profiles, we should expect that these effects should only modify the line profiles weakly, especially for high critical density forbidden lines such as the [\ion{Ne}{II}] $12.81~\mu\mathrm{m}$ and [\ion{O}{I}] $6300$ \r{A} for which the emission mainly traces regions of higher density near the wind base.
Moreover, due to the small volumes concerned, the line profiles are not strongly dependent on the inner radius of the grid - this implies that the effect of gravity in preventing launch at $r < 0.1 r_G$ will not be significant. 

On the other hand, changing from the isothermal to the non-isothermal streamlines may have a stronger impact on the results.
We find the Mach numbers at the base are only a little smaller (Table \ref{tab:Mb_niT}) and there was relatively little difference in either these or the streamline morphology (Figure \ref{fig:streamline_comparison_analytic}).
However, the dominant factor setting the shift of the lines is the sound speed \citep{Ballabio_2020}, thus the fact that the sound speed would vary significantly with strong temperature gradients should likely be taken into account in future work. The development of self-similar models for the non-isothermal case makes this significantly more tractable.

\subsection{Implications for Dust Transport}
Knowing the launch velocity of the wind is a crucial part in understanding which grains may be lost to the wind since the critical size of dust grain for which a reasonable flux reaches the wind is that which has a positive upwards advection velocity in the disc due to the gas supplying the wind \citep{Hutchison_2021,Booth_2021}; this size is proportional to the wind launch velocity if the disc temperature, height of ionisation front and density at the wind base are held constant \citep{Booth_2021}.
However given the rather narrow range of Mach numbers that result for the most reasonable models discussed in this paper, and the robust predictions made by self-similar models \citep[as were applied by][]{Hutchison_2021}, the effect of the wind kinematics on the entrained grain size is largely subdominant compared with uncertainties in the density at the wind base.
However where $b>2$ in the outer disc, no wind is launched (and consequently no dust is removed). Instead we see streamlines return towards the base may which have implications for dust rain-out \citep{Franz_2020,Hutchison_2021}. Since the outward radial velocity of the re-entering gas exceeds the local escape velocity, it is not clear whether any dust returning to the mid-plane would be retained in the disc and we reserve further investigation for future work.

\section{Conclusions}
\label{sec:conclusions}
In this work we have extended the previous studies of self-similar solutions for thermal disc winds by \citet{Clarke_2016} so as to derive a more general set of scale free wind solutions.
Specifically we relax the assumptions of isothermal gas and perpendicular launch from the disc mid-plane and derive solutions for generalised launch geometry and power law temperature profiles.
We validate these solutions using hydrodynamic simulations and furthermore use hydrodynamic simulations to explore non-scale free conditions, such as imposition of a disjoint power law for the wind base density profile and the inclusion of gravitational and centrifugal forces.

We have analysed these models principally in terms of the streamline morphology and the Mach numbers with which the winds are launched (which control the mass-loss rates).
In doing so we have shown that self-similar solutions have widespread and general applicability to describing thermal winds launched from discs at reasonably large radii (beyond the gravitational radius). This is appropriate for protoplanetary discs where outflows consistent with thermal winds are seen to originate in the outer disc \citep{Pascucci_2020}.
This is important for works which seek to apply self-similar models, for example to interpret line spectra \citep{Ballabio_2020} or study dust transport \citep{Hutchison_2021}.
In particular, we find that:

\begin{enumerate}
    \item Scale free temperature profiles, including radial temperature gradients, still permit self-similar solutions which have a constant Mach number at the base, the value of which depends on the detail of the profile. However, for temperature scaling as the inverse root of the (spherical or cylindrical) radius, the Mach number is decreased by only around $10-20$ per cent compared with the isothermal case.
    \item Instead, the parameter which most strongly influences the launch velocity of the winds is the elevation of the wind base to the midplane. The higher the winds originate, the more rapidly they must curve and so the more slowly they are launched.
    \item Scale free hydrodynamic simulations adopt the maximum Mach number at the base for which the solution avoids any singularities in the fluid equations \citep{Clarke_2016} even when the winds are launched from elevated bases, non-perpendicularly to their base or in the presence of temperature gradients.
    \item This preference for a maximal launch Mach number may be explained by the fact that solutions with lower Mach numbers do not completely fill the computational domain. In such a scenario, the region near the z-axis would thus be inaccessible to the wind and provide no resistance to the pressure in the wind region; the streamlines would spread out to fill it, allowing the wind to launch faster. However, if a reflecting boundary, representing some other constraint on the wind, is placed at lower latitude, then simulations adopt a lower launch velocity commensurate with a self-similar solution which asymptotes to the angle set by the reflecting boundary.
    \item When gravity and centrifugal force are included, the streamlines and $\mathcal{M}_{\rm b}$ predicted by self-similar models remain a good approximation to the true streamlines, particularly at large radii or when the radius of curvature is small. 
    Introducing temperature gradients tends to increase the accuracy because these solutions have a smaller radius of curvature and therefore pressure plays a more important role compared to gravity and centrifugal forces.
    \item The predictions of self-similar winds can also be used to describe the launch velocities of density profiles which are double power laws. The velocities vary smoothly between the values appropriate to the density gradient in each limit of the profile, with the launch velocity from inner disc unaffected by the changes in density at larger radius.
    \item Density gradients that are steeper than $r^{-2}$ do not have a valid self-similar wind solution, though may give rise to non-self-similar outflows. 
    However, we find that when the density gradient steepens beyond this point in the outer disc, an outflow is largely prevented: the region of wind launching extends by no more than a factor $\sim 2$ beyond the radius where the base profile attains this limiting gradient. Instead, streamlines originating from the inner disc curl down towards the wind base at radii beyond the transition radius. This material is however unbound and so does not provide a return flow of material into the inner disc.
    \item Recent models of protoplanetary disc photoevaporation such as those by \citet{Wang_2017,Picogna_2019} are broadly consistent with our findings. This illustrates the value in using self-similar models to capture key behaviours of winds.
\end{enumerate}

\section*{Acknowledgements}
We thank the reviewer for a detailed and considered review which helped us to clarify our communication of our results and the underlying physics.
We are grateful to Raphael Franz and Giovanni Picogna for providing some useful insights regarding the comparison to hydrodynamic simulations.
ADS thanks the Science and Technology Facilities Council (STFC) for a Ph.D. studentship and CJC acknowledges support from the STFC consolidated grant ST/S000623/1.
This work has also been supported by the European Union's Horizon 2020 research and innovation programme PEVAP (grant agreement No. 853022) and also under the Marie Sklodowska-Curie grant agreement No 823823 (DUSTBUSTERS). This work was performed using resources provided by the Cambridge Service for Data Driven Discovery (CSD3) operated by the University of Cambridge Research Computing Service (www.csd3.cam.ac.uk), provided by Dell EMC and Intel using Tier-2 funding from the Engineering and Physical Sciences Research Council (capital grant EP/P020259/1), and DiRAC funding from the Science and Technology Facilities Council (www.dirac.ac.uk).

\section*{Data Availability}
The code used to calculate the maximum launch Mach number and self-similar streamline structure for user specified values of $b$, $\phi_{\rm b}$, $\chi_{\rm b}$ and $\tau$ (spherical and cylindrical cases) is available from the lead author's GitHub: \href{https://github.com/AndrewSellek/SelfSimilarThermalWinds}{https://github.com/AndrewSellek/SelfSimilarThermalWinds} or in a Zenodo release: 
\href{https://doi.org/10.5281/zenodo.4579868}{https://doi.org/10.5281/zenodo.4579868}.
Also included are the table from which Figure \ref{fig:Mb_contours} was generated and the streamline solution files used to plot the self-similar streamlines in Figures \ref{fig:streamline_comparison_analytic}, \ref{fig:ss_geometry_streamlines}, \ref{fig:niT_streamlines}, \ref{fig:nss_p36_streamlines}, \ref{fig:full_streamlines} and \ref{fig:b150_b250_streamlines} and the line profiles in Figure \ref{fig:line_profiles}.




\bibliographystyle{mnras}
\bibliography{biblio} 



\appendix
\section{Derivation of Self-Similar Solution for Scale-Free Non-Isothermal Case}
\label{sec:a1}

Following on from the discussion in Section \ref{sec:mathematical_extension}, we provide in more detail in our derivation of the differential equations governing a self-similar streamline in a more general non-isothermal case.
Throughout, primes represent differentiation with respect to $\tilde{z}$.

In dimensionless form, the equations of motion \ref{eq:eom_perp} and \ref{eq:Bernoulli_general_P} are
\begin{equation}
    \frac{\mathcal{M}_{\rm b}^2 \tilde{u}^2}{\tilde{R}_{\rm eff}} = \frac{r_{\rm b}}{\rho_{\rm b} c_{S, \rm b}^2 \tilde{\rho}} \hat{\vec{l}} \cdot \nabla P
    ,
    \label{eq:eom_perp_ss}
\end{equation}
\begin{equation}
    \mathcal{M}_{\rm b}^2 \tilde{u} \frac{{\rm d}\tilde{u}}{{\rm d}\tilde{s}} + \frac{1}{\rho c_{S, \rm b}^2} \frac{\partial P}{\partial \tilde{s}} = 0
    .
    \label{eq:Bernoulli_general_ss}
\end{equation}
Since the second term in equation \ref{eq:Bernoulli_general_ss} is independent of streamline, then $\mathcal{M}_{\rm b}$ must be to ensure the first term is too.

The pressure gradient term can be resolved in terms of changes between streamlines ($r_{\rm b}$) and along the streamlines ($\tilde{s}$)
\begin{align}
    \frac{1}{\rho} \hat{l} \cdot \nabla P
    &= \frac{1}{\rho} \frac{{\rm d}P}{{\rm d}l} = -\frac{1}{\rho} \left(
    \frac{1}{r_{\rm b}} \cot(\chi) \left(\frac{\partial P}{\partial \tilde{s}}\right)_{r_{\rm b}}
    + \frac{1}{\tilde{r} \sin(\chi)} \left(\frac{\partial P}{\partial r_{\rm b}}\right)_{\tilde{s}}
    \right)
    ,
    \label{eq:Pgradient}
\end{align}
where $dl$ is an infinitesimal step in the perpendicular direction $\hat{l}$.

The first term in equation \ref{eq:Pgradient} (due to variation between points on a given streamline), is evaluated by eliminating $\left(\frac{\partial P}{\partial \tilde{s}}\right)_{r_{\rm b}}$ using equation \ref{eq:Bernoulli_general_ss}.
The second term in equation \ref{eq:Pgradient} (due to variation between streamlines) is evaluated by expanding the pressure as $\frac{{\rm d}P}{\rho} = c_S^2 {\rm d} \ln(\rho) + {\rm d}c_S^2$:
\begin{align}
    -\frac{1}{\rho} \frac{1}{\tilde{r} \sin(\chi)} \left(\frac{\partial P}{\partial r_{\rm b}}\right)_{\tilde{s}} =
    \frac{c_{S, {\rm b}}^2}{r_{\rm b}} \frac{b + \tau}{\tilde{r}  \sin(\chi)} \tilde{c}_S^2.
    \label{eq:dPdrb}
\end{align}

Thus combining equations \ref{eq:eom_perp_ss}, \ref{eq:Pgradient}, \ref{eq:dPdrb} and \ref{eq:Bernoulli_general_ss}, we get an equation relating the streamline curvature and the velocity gradients:
\begin{equation}
    \frac{\mathcal{M}_{\rm b}^2 \tilde{u}^2}{\tilde{R}_{\rm eff}}
    = \frac{b+\tau}{\tilde{r} \sin(\chi)} \tilde{c}_{S}^2 - \mathcal{M}_{\rm b}^2 \tilde{u} \frac{{\rm d}\tilde{u}}{{\rm d}\tilde{z}} \cot(\chi) \sin(\theta) = 0
    \label{eq:eom}
\end{equation}
In the case of $\tau=0$ and $\mathcal{C}=1$, the first term recovers the first term of equation 11 of \citet{Clarke_2016}. Moreover, so long as the resulting pressure is scale free, the form of the second term does not depend on the exact temperature structure.

Our generalised geometry makes a simple fundamental change; the area of a streamline bundle normalised to its base becomes
\begin{equation}
    \tilde{A} = \tilde{r}^2 \frac{\sin(\chi)\cos(\phi)}{\sin(\chi_{\rm b})\cos(\phi_{\rm b})},
    \label{eq:Atilde}
\end{equation}
which we use to relate the area and shape of the streamlines, in particular the radius of curvature $\tilde{R}_{\rm eff}$, through the following equations
\begin{align}
    \tilde{R}''
    &= \frac{(1+\tilde{R}'^2) (\tilde{R} - \tilde{z} \tilde{R}') \tilde{R}'}{\tilde{R}(\tilde{z}+\tilde{R}\tilde{R}')}
    - \frac{(1+\tilde{R}'^2)^{3/2}}{\tilde{R}(\tilde{z}+\tilde{R}\tilde{R}')} \tilde{A}' \cos(\phi_{\rm b})\sin(\chi_{\rm b}),
    \label{eq:R_dlbprime} \\
    \frac{1}{\tilde{R}_{\rm eff}} &= \frac{\tilde{R}''}{(1+\tilde{R}'^2)^{0.5}},
    \label{eq:Reff}
\end{align}
in which $\tilde{A}'$ is given by our nozzle equation \ref{eq:nozzle_ss},
\begin{equation}
    \frac{{\rm d}\ln(\tilde{A})}{{\rm d} \tilde{z}} = \left( \mathcal{M}_{\rm b}^2 \frac{\tilde{u}^2}{\tilde{c}_s^2} -1 \right) \frac{{\rm d}\ln(\tilde{u})}{{\rm d} \tilde{z}} + \frac{{\rm d}\ln(\tilde{c}_S^2)}{{\rm d} \tilde{z}}
    .
    \label{eq:nozzle_ss}
\end{equation}
If we expand ${\rm d}\tilde{c}_S^2(\tilde{r},\phi) = \tilde{r}^{-\tau} \mathcal{C}(\phi)$ using the chain rule,
\begin{equation}
    \frac{{\rm d} \tilde{c}_S^2}{{\rm d} \ln(\tilde{r})}
    = \tilde{c}_S^2 \left( - \tau  + \frac{\partial \ln(\mathcal{C})}{\partial \phi} \tan(\chi) \right)
    ,
\end{equation}
we obtain a final relation between the variation of the area, the velocity gradient and the imposed self-similar temperature structure\footnote{Note that when evaluated in this way, $\tilde{A}' \propto \tilde{A}$, such that the $\tilde{A}' \cos(\phi_{\rm b}) \sin(\chi_{\rm b})$ term in equation \ref{eq:R_dlbprime} is actually independent of $\phi_{\rm b}$ and $\chi_{\rm b}$, which therefore do not appear in our final expressions.}:
\begin{equation}
    \frac{\tilde{A}'}{\tilde{A}}
    = \left(\mathcal{M}_{\rm b}^2 \frac{\tilde{u}^2}{\tilde{c}_S^2} - 1 \right) \frac{{\rm d}\ln(\tilde{u})}{{\rm d}\tilde{z}}
    + \left( - \tau
    + \frac{\partial \ln(\mathcal{C})}{\partial \phi} \tan(\chi) \right)
    \frac{\cos(\chi)}{\tilde{r}\sin(\theta)}
    \label{eq:dlnA_general}.
\end{equation}

Combining Equations \ref{eq:eom}-\ref{eq:Reff} and \ref{eq:dlnA_general} gives Equations \ref{eq:odes_0}-\ref{eq:odes_g2}.

\section{Equations for Power Law Temperature Profiles}
\label{sec:a2}
\subsection{Spherical Power Law}

When $T \propto r^{-\tau}$, $\mathcal{C}(\phi) = 1$, such that we have:
\begin{align}
    f_1 &= - \mathcal{M}_b^2 \tilde{u} \left(\mathcal{M}_b^2 \tilde{u}^2 (\tilde{R}^2 + \tilde{z}^2)^{\tau/2} -1\right)
    \frac{(\tilde{R}-\tilde{z}\tilde{R}')}{(1+\tilde{R}'^2)^{1/2}(\tilde{R}\tilde{R}'+\tilde{z})} \label{eq:f1_nit}\\
    g_1 &= (b+\tau) \frac{(1+\tilde{R}'^2)^{1/2}  }{(\tilde{R}-\tilde{z}\tilde{R}')(\tilde{R}^2+\tilde{z}^2)^{\tau/2}}
    \label{eq:g1_niT} \\
    g_2 &= - \mathcal{M}_b^2 \tilde{u}^2
    \frac{(\tilde{R}-\tilde{z}\tilde{R}')}{(1+\tilde{R}'^2)^{1/2}}
    \left(\frac{\tilde{R}'}{\tilde{R}(\tilde{R}\tilde{R}'+\tilde{z})}
    + \frac{\tau}{(\tilde{R}^2+\tilde{z}^2)}
    \right)
    \label{eq:g2_nit}
    .
\end{align}

\subsection{Cylindrical Power Law}

When $T \propto R^{-\tau}$, $\mathcal{C} = \left( \frac{\cos(\phi)}{\cos(\phi_{\rm b})} \right)^{-\tau}$ in which case $\frac{\partial \ln(\mathcal{C})}{\partial \phi} = \tau \tan(\phi)$, such that we have:
\begin{align}
    f_1 &= - \mathcal{M}_b^2 \tilde{u} \left(\mathcal{M}_b^2 \tilde{u}^2 \frac{\tilde{R}^{\tau}}{\cos^\tau(\phi_{\rm b})} -1\right)
    \frac{(\tilde{R}-\tilde{z}\tilde{R}')}{(1+\tilde{R}'^2)^{1/2}(\tilde{R}\tilde{R}'+\tilde{z})} \label{eq:f1_nit_cyl}\\
    g_1 &= (b+\tau) \frac{(1+\tilde{R}'^2)^{1/2}  }{(\tilde{R}-\tilde{z}\tilde{R}')\tilde{R}^\tau}  \cos^\tau(\phi_{\rm b})
    \label{eq:g1_nit_cyl} \\
\begin{split}
    g_2 &= - \mathcal{M}_b^2 \tilde{u}^2
    \frac{(\tilde{R}-\tilde{z}\tilde{R}')}{(1+\tilde{R}'^2)^{1/2}}
    \\
    &\left(\frac{\tilde{R}'}{\tilde{R}(\tilde{R}\tilde{R}'+\tilde{z})}
    + \frac{\tau}{(\tilde{R}^2+\tilde{z}^2)}
    \left[1-\frac{\tilde{z}(\tilde{R}-\tilde{z}\tilde{R}')}{\tilde{R}(\tilde{R}\tilde{R}'+\tilde{z})}\right]\right)
    \label{eq:g2_nit_cyl}
    .
\end{split}
\end{align}

\bsp	
\label{lastpage}
\end{document}